\documentclass[aps,prb,twocolumn,showpacs,floatfix]{revtex4-1}
\usepackage{amssymb,amsmath} 
\usepackage{graphicx}
\usepackage{dcolumn}
\usepackage{bm}
\usepackage{epstopdf} 
\usepackage{color}
\usepackage[colorlinks,bookmarks=false,citecolor=blue,linkcolor=blue,urlcolor=blue]{hyperref}
\def \S{{\bf S}}

\def \k{{\bf k}}
\def \Q{{\bf Q}}
\def\G{{\bf G}}
\def\r{{\bf r}}
\def\ahat{\hat{a}}
\def\bhat{\hat{b}}
\def \xhat{{\hat{x}}}
\def \yhat{{\hat{y}}}
\def\shat{\hat{s}}
\def\that{\hat{t}}
\def\qhat{\hat{q}}
\def\hhat{\hat{h}}
\def\gammahat{\hat{\gamma}}

\def\sbar{\bar{s}}
\def\udel{\underline{\delta}}
\def\calN{\mathcal{N}}
\def\calA{\mathcal{A}}
\def\calB{\mathcal{B}}
\def\one{\mathbf{1}}
\def\onebar{{\bar{\mathbf{1}}}}
\def\zero{\mathbf{0}}
\begin{document}
\title{Bond-operators and triplon analysis for spin-$S$ dimer antiferromagnets}
\author{Brijesh Kumar}
\email{bkumar@mail.jnu.ac.in}
\affiliation{School of Physical Sciences, Jawaharlal Nehru University, New Delhi 110067, India.}
\date{September 21, 2009 }
\begin{abstract}
The mean-field triplon analysis is developed for spin-$S$ quantum antiferromagnets with dimerized ground states. For the spin-1/2 case, it reduces to the well known bond-operator mean-field theory. It is applied to a coupled-dimer model on square lattice, and to a model on honeycomb lattice with spontaneous dimerization in the ground state. Different phases in the ground state are investigated as a function of spin. It is found that under suitable conditions (such as strong frustration)  a quantum ground state (dimerized singlet phase in the present study) can survive even in the limit $S\rightarrow \infty$. Two quick extensions of this representation are also presented. In one case, it is extended to include the quintet states. In another, a similar representation is worked out on a square plaquette. A convenient procedure for evaluating the total-spin eigenstates for a pair of quantum spins is presented in the appendix.
\end{abstract}
\maketitle
\section{Introduction}
\label{sec:introduction}
Antiferromagnetically interacting spins are sensitive to both quantum mechanics and frustration~\cite{Lhuillier,Indrani}. Therefore, the quantum antiferromagnets with frustration can realize interesting non-magnetic ground states such as the dimerized singlet (valence bond) states~\cite{MG, SS, Kageyama}, plaquette singlets~\cite{CaV4O9, CaV4O9_Troyer},  spin-liquids~\cite{Kagome_Material, Mila,kappa_salt} etc., apart from having an antiferromagnetically ordered ground state (say, N\'eel type). In the present study, we are concerned with those systems in which the ground state has a dimer order, either spontaneous or given. There is an ever-increasing number of quantum antiferromagnetic (AF) materials which exhibit, or seem to exhibit, dimerization physics at low temperatures~\cite{Kageyama, TlCuCl3, TlCuCl3Nature, BaCuSi2O6,CuFeGeO,Sr3Cr2O8}. Such systems are typically characterized by an energy gap to spin excitations, thereby showing sharp drop in the magnetic susceptibility as the temperature is lowered below a certain temperature characteristic of the interaction between spins. 

We are presently interested in generic theoretical questions concerning the instability of a dimer singlet ground state towards AF ordering as competing interactions in a system are varied. For a spin-1/2 system, such investigations at the simplest level can be conveniently carried out by doing a mean-field triplon analysis with respect to a dimer ground state. A {\em triplon} is a triplet excitation residing on a bond (dimer), and dispersing from one bond to another under the exchange interactions present in the system. An energy gap in the triplon dispersion implies a stable dimer phase, while gaplessness signifies an AF order of some kind (that depends upon the dispersion). The underlying formulation is facilitated by what is called the {\em bond-operator} representation of the spin operators~\cite{sach-bhatt,chub}. This approach has been successfully applied to many different spin-1/2 systems~\cite{gopalan_rice,rkbk,rkdkbk}. Subsequent to this, the bond-operator method has also been developed for spin-1 dimer problems~\cite{spin1_bo1,spin1_bo2,spin1_bo3, spin1_bo4}. Moreover, the triplon analysis has also been suitably extended to the square-plaquette problems (spin-1/2 case)~\cite{plaquette1}. However, no such formulation exists for a general spin-$S$ dimer problem. In the present work, we precisely set out to achieve this objective. That is, to derive the bond-operator representation for spin-$S$ operators (in Sec.~\ref{sec:boS}), and to do the mean-field triplon analysis for some model systems of interest (in Secs.~\ref{sec:coupledCD} and~\ref{sec:honeycomb}). 

Obviously, it is impractical to be working with all the $(2S+1)^2$ states of a spin-S dimer. We therefore restrict our analysis to the subspace of singlet and triplet states only. Apart from simplifying our labor, which it does, it is enough for a primary discussion of the problem. Here, we adopt a simple working philosophy that, for an antiferromagnetic spin-S dimer problem, the triplet excitations are the principal cause of instability (if it occurs) of a singlet phase, as the higher spin excitations are further up in energy and hence irrelevant for an effective low energy description. It is implicit in our discussion that a given system only has exchange interactions. The treatment, however, will have to be extended to include quintet or higher total-spin states, if the single-ion anisotropy effects are present. While it is a real concern, presently we focus only on developing the triplon analysis for dimerized spin-$S$ quantum antiferromagnets. As an interesting byproduct of this exercise, we have also developed a nice and simple method for evaluating the total-spin eigenstates for a pair of spin-$S$ (angular momentum addition) using Schwinger boson representation. Our method resembles that of Schwinger's, but it is different in actual details of the procedure that generates the compound spin eigenstates.~\cite{Schwinger,Mattis} For the benefit of readers, it is presented in detail in the appendix.

Of the two spin-$S$ quantum antiferromagnets, that we apply this mean-field triplon analysis to, the first one in Sec.~\ref{sec:coupledCD} is a coupled columnar-dimers model on square lattice. Different variations of this model for the spin-1/2 (and spin-1) case have been studied extensively for investigating the quantum phase transition from dimer to AF ordered phases~\cite{SinghGelfandHuse, AffleckGelfandSingh, Matsumoto, Ivanov, Imada, Richter_SpinDependence,Wenzel_Janke}. Here, we investigate this transition as a function of $S$, the size of spin. Interestingly, we find that the dimer singlet phase (a quantum mechanical phase) survives even in the so-called classical limit ($S\rightarrow\infty$), under suitable conditions (such as strong frustration). For example, we find the columnar-dimer phase to be stable and present in the limit $S\rightarrow\infty$ in a small range of coupling around $J_2/J_1=0.5$ for the $J_1$-$J_2$ model. While other kinds of gapped phases (such as the Haldane phase) are likely to arise in different regions of the phase diagram (especially for $S\ge 1$)~\cite{Matsumoto}, we can not study them here within a bond-operator mean-field theory. However, the conclusion drawn from the present calculation about the $S\rightarrow\infty$ limit, which is certainly valid in the vicinity of strongly dimerized limit, suggests a generic possibility of this kind, and asks for a rethinking of the classical limit in quantum antiferromagnets.

The second one is a model on honeycomb lattice, recently constructed and shown by the present author to have an exact triply degenerate dimer ground state for a certain value of the interaction parameter, for any spin. In Ref.~\onlinecite{rkdkbk}, we have already presented the mean-field triplon analysis results for the spin-1/2 case of this model. However, we did not know then, how to do it for $S>1/2$. This, in fact, was our original motivation behind developing the triplon analysis for spin-$S$ dimer problems. From the mean-field triplon analysis of this model in Sec.~\ref{sec:honeycomb}, for $S\ge1$, we find the dimer phase giving way to the N\'eel ordered AF phase  as soon as one moves away from the point of exact dimer ground state. It is unlike the spin-1/2 case where the dimer phase was found to survive over a finite range of coupling. 

In Sec.~\ref{sec:extend}, we present two straightforward extensions of the spin-S bond-operator representation. First, we extend it to include the quintet states. Next, we work out a similar representation on a square plaquette, in the restricted space of a plaquette singlet and certain low-lying triplets. Finally, we conclude with a summary.

\section{Bond-operator representation for spin-$S$ operators}
\label{sec:boS}
Consider the Heisenberg exchange interaction,  $\S_1\cdot\S_2$, on a bond. The eigenstates, $\{|j,m_j\rangle \}$, of this problem are such that, $\S_1\cdot\S_2|j,m_j\rangle=[-S(S+1)+\frac{1}{2}j(j+1)]|j,m^{ }_j\rangle$, where $j=0,1,\dots,2S$ is the total-spin quantum number of the two spins. For a given $j$, the eigenvalues $m_j$ of operator $(\S_1+\S_2)_z$ are given by $m^{ }_j=-j,-j+1,\dots, j$. Therefore, the bond eigenstate for a given $j$ is ($2j+1$)-fold degenerate. It is a singlet for $j=0$, triplet for $j=1$, quintet for $j=2$, and so on. We denote the singlet state as $|s\rangle$, and the triplets as $|t^{ }_{m^{ }_1}\rangle $ where $m^{ }_1=0,\pm 1$. The quintet states are denoted as $|q^{ }_{m^{ }_2}\rangle$ where $m^{ }_2=0,\pm 1,\pm 2$. The eigenstates for $j > 2$ may in general be denoted as $|h_{j,m_j}\rangle$. Below we define the bosonic creation operators, $ \shat^\dag, \that^\dag_{m^{ }_1}$, $ \qhat^\dag_{m^{ }_2}$ and $ \hhat^\dag_{j,m_j}$, corresponding to the bond eigenstates. These operators are called bond-operators. 
\begin{subequations}
\begin{eqnarray}
|s\rangle &:=& \shat^\dag|0\rangle \label{eq:sbop}\\
|t^{ }_{m^{ }_1}\rangle &:=& \that^\dag_{m^{ }_1}|0\rangle \label{eq:tbop}\\
|q^{ }_{m^{ }_2}\rangle &:=& \qhat^\dag_{m^{ }_2} |0\rangle \label{eq:qbop}\\
|h^{ }_{j,m_j}\rangle &:=& \hhat^\dag_{j,m_j}|0\rangle \label{eq:hbop}
\end{eqnarray}
\end{subequations}
Here, $|0\rangle$ denotes the vacuum of the bosonic Fock space. The completeness of the bond eigenstates implies the following physical constraint on the bond-operators.
\begin{equation}
\shat^\dag\shat + \that^\dag_{m^{ }_1}\that^{ }_{m^{ }_1}+\qhat^\dag_{m^{ }_2}\qhat^{ }_{m^{ }_2}+\hhat^\dag_{j,m_j}\hhat^{ }_{j,m_j}=1
\label{eq:constraint}
\end{equation}
Here, the repeated indices are summed over. 

In terms of the bond-operators, the exchange Hamiltonian on the bond can be written as:
\begin{eqnarray}
& J\S_1\cdot\S_2 = \nonumber \\
& -JS(S+1)\left[\shat^\dag\shat+\that^\dag_{m^{ }_1}\that^{ }_{m^{ }_1}+\qhat^\dag_{m^{ }_2}\qhat^{ }_{m^{ }_2} +\hhat^\dag_{j,m_j}\hhat^{ }_{j,m_j}\right] \nonumber \\
& +J\left[\that^\dag_{m^{ }_1}\that^{ }_{m^{ }_1} +3\qhat^\dag_{m^{ }_2}\qhat^{ }_{m^{ }_2}+\frac{1}{2} j(j+1) \hhat^\dag_{j,m_j}\hhat^{ }_{j,m_j}\right]. \label{eq:bondH}
\end{eqnarray}
Furthermore, to describe the interaction between the spins of different bonds in the bosonic Fock space, we must know the spin operators in terms of the bond-operators. Below we develop the bond-operator representation for spins, which is a generalization of the well-known bond-operator representation for spin-1/2 operators to the case of arbitrary spin-$S$.

In order to construct the bond-operator representation for the spins, we first find out the explicit  forms of all the eigenstates on a bond. In the appendix to this paper, we have worked out an elegant procedure to write down the total-spin eigenstates for a pair of arbitrary spins. Following this approach, we can write the singlet wavefunction on a bond as:
\begin{equation}
|s\rangle = \frac{1}{\sqrt{2S+1}}\sum_{m=0}^{2S} (-)^m |S-m,-S+m\rangle. \label{eq:singlet_state}
\end{equation}
Here, the state, $|S-m,-S+m\rangle$, denotes a product-state, $|S,S-m\rangle \otimes|S,-S+m\rangle$, of the two spins of a bond. For the derivation of Eq.~(\ref{eq:singlet_state}), refer to Proposition 1 in the appendix, and also see Fig.~\ref{fig:singlet_triplets}. We can similarly write the three triplet states as:
\begin{subequations}
\begin{eqnarray}
|t_1\rangle &=& \frac{1}{\sqrt{{\tt N}_t}}\sum_{m=0}^{2S-1}(-)^m \sqrt{(2S-m)(m+1)} \times \nonumber \\
&& \hspace{2cm}  |S-m,-S+m+1\rangle \label{eq:triplet_plus1}\\
|t_0\rangle &=& \sqrt{\frac{2}{{\tt N}_t}}\sum_{m=0}^{2S}(-)^m (S-m) |S-m,-S+m \rangle \label{eq:triplet_0} \\
|t_{\bar{1}}\rangle &=& \frac{1}{\sqrt{{\tt N}_t}}\sum_{m=0}^{2S-1}(-)^m\sqrt{(2S-m)(m+1)} \times \nonumber \\
&&  \hspace{2cm} |S-m-1,-S+m\rangle \label{eq:triplet_minus1}
\end{eqnarray}
\end{subequations}
where $\bar{1}$ in the above equation denotes $m^{ }_1=-1$ (we will sometime denote negative integers as  integers with a bar), and the normalization, ${\tt N}_t=2S(S+1)(2S+1)/3$. 
Refer to Eqs.~(\ref{eq:1plus1}) to (\ref{eq:1minus1}) and Fig.~\ref{fig:singlet_triplets} for the derivation of the triplet states. Below we also write the quintet states for $m^{ }_2=0$, $1$ and $2$ (refer to Fig.~\ref{fig:quintets} and related discussion for details).
\begin{subequations}
\begin{eqnarray}
|q^{ }_2\rangle &=& \frac{1}{\sqrt{{\tt N}_q}}\sum_{m=0}^{2S-2}(-)^m \sqrt{(2S-m)(2S-m-1)} \nonumber \\
&&  \times \sqrt{(m+1)(m+2)} \, |S-m,-S+m+2\rangle \label{eq:quintet_plus2}\\
|q^{ }_1\rangle &=& \frac{1}{\sqrt{{\tt N}_q}}\sum_{m=0}^{2S-1}(-)^m (2S-2m-1)\sqrt{2S-m}  \nonumber\\
&& \hspace{1cm} \times \sqrt{m+1} \, |S-m,-S+m+1 \rangle 
\label{eq:quintet_plus1} \\
|q^{ }_{0}\rangle &=&  \sqrt{\frac{2}{3{\tt N}_q}}\sum_{m=0}^{2S}(-)^m [3(S-m)^2-S(S+1)]  \nonumber \\
&& \hspace{2cm} \times |S-m,-S+m\rangle \label{eq:quintet_0}
\end{eqnarray}
\end{subequations}
Here, ${\tt N}_q=2S(S+1)(2S-1)(2S+1)(2S+3)/15$. The state $|q^{ }_{\bar{1}} \rangle$ can be generated from $|q^{ }_1\rangle$ by changing $|S-m,-S+m+1\rangle $ to $|S-m-1,-S+m\rangle$ (that is, $\S_{1z} \leftrightarrow -\S_{2z}$). Similarly, for $|q^{ }_{\bar{2}}\rangle$. We can also evaluate the states for higher $j$ values following the Propositions 5 and $5^*$ in the appendix.

As emphasized earlier, the present discussion will be restricted to the subspace of singlet and triplet states only. By computing the matrix elements of $\S_1$ and $\S_2$ in this restricted subspace, we derive the following bond-operator representation for spin-$S$ operators.
\begin{subequations}
\begin{eqnarray}
{\bf S}_{1\alpha} &\approx& \sqrt{\frac{S(S+1)}{3}}\left(\shat^\dag\that^{ }_\alpha + \that^\dag_\alpha\shat\right)-\frac{i}{2}\epsilon_{\alpha\beta\gamma}\that_\beta^\dag\that^{ }_\gamma \label{eq:bo_s1}\\
{\bf S}_{2\alpha} &\approx& -\sqrt{\frac{S(S+1)}{3}}\left(\shat^\dag\that^{ }_\alpha + \that^\dag_\alpha\shat\right)-\frac{i}{2}\epsilon_{\alpha\beta\gamma}\that_\beta^\dag\that^{ }_\gamma
\label{eq:bo_s2}
\end{eqnarray}
\end{subequations}
Here, $\alpha=x,y,z$ (for three components of the spin operators), and the same for $\beta$ and $\gamma$. The $\epsilon_{\alpha\beta\gamma}$ denotes the totally antisymmetric tensor. Moreover, $\that^\dag_x = \frac{1}{\sqrt{2}}(\that^\dag_{\bar{1}}-\that^\dag_1)$, $\that^\dag_y = \frac{i}{\sqrt{2}}(\that^\dag_{\bar{1}}+\that^\dag_1)$ and $\that^\dag_z= \that^\dag_0$. Since it is convenient to write the bond-operator representation, Eqs.~(\ref{eq:bo_s1}) and (\ref{eq:bo_s2}), in terms of $\that^\dag_\alpha$, we also write the constraint, Eq.~(\ref{eq:constraint}), and  the bond Hamiltonian, Eq.~(\ref{eq:bondH}), using the same, replacing $\that^\dag_{m^{ }_1}\that^{ }_{m^{ }_1}$ by $\that^\dag_\alpha \that^{ }_\alpha$.

While our representation is valid for arbitrary $S$, it is obviously approximate for $S\ge 1 $, because it is constructed in a restricted subspace, ignoring the contributions from $j\ge 2$ states. However, it is exact for $S=1/2$, and correctly reproduces the known representation~\cite{sach-bhatt}. As briefly discussed in the introduction, for doing a simple stability analysis of the dimer phase of a spin-$S$ quantum antiferromagnet, it would suffice to know the dynamics of triplet excitations, except when it may be necessary to consider higher spin states. 

In the following sections, we do mean-field triplon analysis of two different spin-$S$ models using this bond-operator representation. The first one  is a model of  coupled columnar-dimers on square lattice. It reduces to many different models of interest such as $J_1$-$J_2$ model, the coupled ladders and so on. The second model is defined on the honeycomb lattice. It admits an exact dimer ground state for arbitrary $S$, and is expected to undergo a transition to the N\'eel ordered phase away from the exactly solvable case. 
\section{Coupled dimers on square lattice}
\label{sec:coupledCD}
\begin{figure}[t]
   \centering
   \includegraphics[width=7cm]{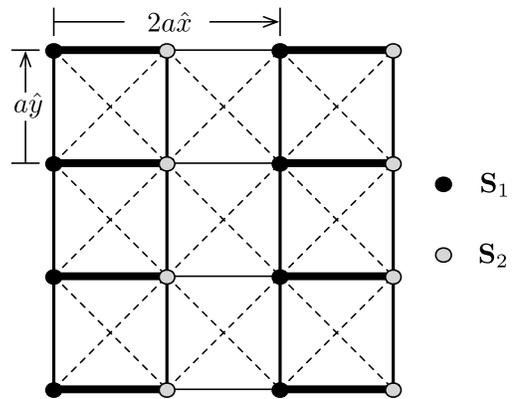}
   \caption{Quantum spin-$S$ coupled dimer model. As in Eq.~\ref{eq:model1}, the exchange interactions are: thick bonds=$J$, thin horizontal lines=$\eta_x J$, not-so-thin vertical lines=$\eta_y J$, and thin dashed lines=$\eta^\prime J$, where $J$, $\eta_x$, $ \eta_y$ and  $\eta^\prime>0$. Also shown are the primitive translations, and the spin labeling on a dimer.}
   \label{fig:model1}
\end{figure}
Consider a spin-$S$ quantum antiferromagnet of interacting dimers on square lattice as shown in Fig.~\ref{fig:model1}. The arrangement of dimers is taken to be columnar because it occurs in the disordered ground state of the $J_1$-$J_2$ model, which is a special case of the present model.
However, one may also consider other arrangements, if one wants. The Hamiltonian of this model is written below.
\begin{eqnarray}
H^{ }_{\tt I} &=& J \sum_{\r} \Big[ \S_{\r,1} \cdot \S_{\r,2}  +  \eta_x \S_{\r,2} \cdot \S_{\r+2a\xhat,1} \nonumber \\
&& + \eta_y \left( \S_{\r,1} \cdot \S_{\r+a\yhat,1}+ \S_{\r,2} \cdot \S_{\r+a\yhat,2}\right) \nonumber\\
&& +\eta^\prime\sum_{\udel}^{\pm a\yhat}\left( \S_{\r,1} \cdot \S_{\r+\udel,2}+\S_{\r,2} \cdot \S_{\r+2a\xhat+\udel,1} \right) \Big]
\label{eq:model1}
\end{eqnarray}
Here, $1$ and $2$ denote the two spins of a dimer; $\r$ denotes the position of a dimer, and is summed over all dimers; $J$ is the intra-dimer antiferromagnetic exchange; various inter-dimer spin-exchange interactions (relative to $J$) are given by $0\le \eta_x, \eta_y, \eta^\prime\le 1$. Refer to Fig.~\ref{fig:model1} for clarification. Physically, the single-ion anisotropy is also expected to be present for $S \ge 1$ (except when the ground state of the ion is an orbital singlet). However, presently we work without such anisotropy effects. Our objective here is to apply the bond-operator representation of the previous section to study the stability of a dimer phase in a simple model quantum antiferromagnet, and not study any particular physical system. The model given by $H_{\tt I}$ reduces to the following simpler models of interest for certain special choices of the interactions.
\begin{enumerate}
\item $\eta^\prime=0$. It presents an {\em unfrustrated}  model which interpolates between coupled dimer chains and coupled ladders, passing through a square lattice model at $\eta_x=\eta_y$. For $\eta_x\simeq 0$,  it describes a set of decoupled (or weakly coupled) two-leg ladders, which is physically the case of a family of ladder compounds Na$_2$T$_2$(C$_2$O$_4$)$_3$(H$_2$O)$_2$, where T$=$Ni, Co, Fe and Mn in the increasing order of spin from $S=1$ to $5/2$. As noted earlier, one must also consider the single-ion anisotropy in real materials, which has been ignored presently. A realistic calculation for this family of ladder compounds will be discussed elsewhere. For $\eta_y=0$, it describes a set of decoupled dimerized spin chains.
\item $\eta^\prime \neq 0$ and $\eta_x=\eta_y=\eta$. It is a frustrated model of coupled columnar dimers on square lattice which reduces to the well-known $J_1$-$J_2$ model when $\eta=1$.
\end{enumerate}
Below the two cases are studied within a triplon mean-field theory of $H_{\tt I}$. This is however not an exhaustive discussion of the problem, as it does not address other kinds of gapped phases that may arise for larger values of $S$. In any case, let us see what we learn from this simple  stability analysis of the dimer phase.
\subsection{Mean-field triplon analysis}
\label{subsec:mft}
Imagine a special limit of $H_{\tt I}$ in which all couplings, except the intra-dimer exchange $J$, were zero. Then, the spins on each thick bond (dimer) in Fig.~\ref{fig:model1} would form an exact singlet in the ground state. Moreover, the elementary excitation in this case would correspond to creating a triplet on it. Since a triplet state on a bond costs an extra energy $J$, there is an energy gap to such excitations. Besides, these triplet excitations are localized because of the absence of inter-dimer interactions in this special case of independent dimers. This limit presents an idealized version of what is called a {\em spin-gapped} dimer phase in quantum antiferromagnets. In general,  the inter-dimer spin interactions are non-zero, and the triplet excitations disperse, thereby lowering the spin-gap. A dispersing bond-triplet is often called a {\em triplon}.  As long as the triplon spin-gap is non-zero, the dimer phase is stable against these excitations, and such a ground state will have zero magnetic moment. For some values of the competing  interactions in a problem, the spin-gap may however close. This marks the onset of a quantum phase transition from the gapped dimer phase to a gapless ordered AF phase. We study such quantum phase transitions in $H_{\tt I}$ within a simple mean-field theory using bond operators.

The key steps of a mean-field triplon analysis are the following. First, identify a configuration of the singlet forming dimers as expected in the ground state. In the present case, by construction, the preferred dimers are the tick bonds in Fig.~\ref{fig:model1}. Using the bond-operator representation for the spins on each dimer, rewrite the spin Hamiltonian in terms of the bond operators, including the constraint by means of a Lagrange multiplier. Now replace the singlet bond-operators, $\shat$ and $\shat^\dag$, on every dimer by a mean-field, $\sbar$. This results in a model of interacting triplons (on a mean-field singlet background given by $\sbar$). To make the problem tractable, ignore the trilpon-triplon interaction (similar to the spin-wave analysis). The last two steps essentially amount to writing the spins on a dimer as: \( \S_{1\alpha} =-\S_{2\alpha} \approx \sbar \sqrt{S(S+1)/3} (\that^\dag_\alpha +\that^{ }_\alpha) \), and replacing $\shat^\dag\shat$ by $\sbar^2$ in Eqs.~(\ref{eq:constraint}) and (\ref{eq:bondH}). As an additional simplification, we satisfy the bond-operator constraint only globally. This gives us a bilinear Hamiltonian of triplons which can be studied fairly straightforwardly. In the present formulation, the quintets and higher bond eignestates for $S\ge 1$ are dispersion-less higher energy excitations, and will play no role in determining the ground state properties.

Applying the above prescription to $H_{\tt I}$ gives the following mean-field triplon Hamiltonian. 
\begin{eqnarray}
H^{ }_{{\tt I},mf} = N_d\left[J-JS(S+1)-\frac{5}{2}\lambda +\sbar^2 (\lambda-J)\right] + \nonumber\\ \frac{1}{2}\sum_{\k,\alpha}\Bigg\{[\lambda-\sbar^2 S(S+1) \xi_{\k} ]\left(\that^\dag_{\k\alpha}\that^{ }_{\k\alpha} + \that^{ }_{-\k\alpha}\that^\dag_{-\k\alpha}\right) \nonumber \\
- \sbar^2 S(S+1)\xi_{\k} \left(\that^\dag_{\k\alpha}\that^\dag_{-\k\alpha} + \that^{ }_{-\k\alpha}\that^{ }_{\k\alpha}\right) \Bigg\} +\nonumber \\
\sum_{\k}\sum^{j=2S}_{j\ge 2}\sum_{m_j}\left[\lambda+\frac{J}{2}j(j+1)-J\right] \hhat^\dag_{\k,j,m_j} \hhat^{ }_{\k,j,m_j} \label{eq:H1mf}
\end{eqnarray}
Here, $N_d$ is the number of dimers, and $\xi_{\k} = 2J\epsilon_\k/3$. 
Moreover, \(\epsilon_\k = \eta_x\cos{(2k_xa)}+2(\eta^\prime-\eta_y)\cos{(k_ya)}+2\eta^\prime\cos{(2k_xa)}\cos{(k_ya)}\). 

The triplon part of the mean-field Hamiltonian, $H^{ }_{{\tt I},mf}$, can be diagonalized using Bogoliubov transformation. Define the triplon quasi-particle operators, $\gammahat^{ }_{\k\alpha}$, such that \( \that^{ }_{\k\alpha} = \cosh{\theta_{\k\alpha}} \gammahat^{ }_{\k\alpha} +\sinh{\theta_{\k\alpha}}\gammahat^\dag_{-\k\alpha}\), and demand that the triplon terms in Eq.~(\ref{eq:H1mf}) be diagonal in $\gammahat^{ }_{\k\alpha}$. This is achieved for \(\tanh{2\theta_{\k\alpha}} =\sbar^2 S(S+1) \xi_{\k}/[\lambda-\sbar^2 S(S+1) \xi_{\k}]\), giving the following diagonal mean-field Hamiltonian.
\begin{eqnarray}
H^{ }_{{\tt I},mf} = e_0 N_d + \sum_{\k,\alpha}E_{\k}\left(\gammahat^\dag_{\k\alpha}\gammahat^{ }_{\k\alpha} +\frac{1}{2}\right)+\nonumber \\
\sum_{\k}\sum^{j=2S}_{j\ge 2}\sum_{m_j}\left[\lambda+\frac{J}{2}j(j+1)-J\right] \hhat^\dag_{\k,j,m_j} \hhat^{ }_{\k,j,m_j} \label{eq:diagonalH1mf}
\end{eqnarray}
Here, $e_0 = J-JS(S+1)-(5\lambda/2) +\sbar^2(\lambda-J)$, and $E_{\k} = \sqrt{\lambda[\lambda-2\sbar^2 S(S+1)\xi_{\k}]}$ is the triplon dispersion. The ground state in this mean-field theory is given by the vacuum of the triplon quasi-particles, and of the excitations for $j\ge 2$. The ground state energy per dimer, $e_g$, of the $H^{ }_{{\tt I},mf}$ is given by
\begin{equation}
e_g[\lambda,\sbar^2] = e_0 + \frac{3}{2 N_d} \sum_{\k} E_{\k}. \label{eq:eg1}
\end{equation}

Minimizing $e_g$ with respect to $\lambda$ and $\sbar^2$ 
gives the following mean-field equations.
\begin{subequations}
\begin{eqnarray}
\sbar^2 &=& \frac{5}{2} - \frac{3}{2N_d}\sum_{\k}\frac{\lambda- \sbar^2 S(S+1)\xi_{\k}}{E_{\k}} \label{eq:H1sbar2}\\
\lambda &=& J+\frac{3\lambda S(S+1)}{2N_d}\sum_{\k} \frac{\xi_{\k}}{E_{\k}} \label{eq:H1lambda}
\end{eqnarray}
\end{subequations}
The physical solution corresponds to solving these equations self-consistently for $\sbar^2$ and $\lambda$. This we will do separately for different cases of the model. One can calculate the spin-gap, and also the magnetic moment in the ordered phase, by solving these equations in the entire parameter space. The staggered magnetic moment in the ordered AF phase is given by $M_s=2\sbar\sqrt{S(S+1)n_c/3}$, where $n_c$ is the triplon condensate density in the gapless phase~\cite{rkbk}. Presently, we only compute the phase boundaries between the dimer and the magnetically ordered phases. This is done by tracing the closing of the triplon gap. The wavevector $\Q$, at which the bottom of the dispersion touches zero ($E_\Q =0$), decides the magnetic order in the AF phase. The vanishing triplon gap also fixes $\lambda$ as \(\lambda^*=2\sbar^2 S(S+1) \xi_{\Q}$. After a few steps of algebra on Eqs.~(\ref{eq:H1sbar2}) and (\ref{eq:H1lambda}), we get the following equation for the phase boundary between the columnar dimer phase and the $\Q$-ordered antiferromagnetic phase in the space of coupling parameters.  
\begin{equation} 
\epsilon_{\Q}\left[5-\frac{3}{N_d}\sum_{\k}\sqrt{\frac{\epsilon_\Q}{\epsilon_\Q - \epsilon_\k}} \right] = \frac{1.5}{S(S+1)} 
\label{eq:H1QPBoundary}
\end{equation}
In the above equation, the spin appears as $S(S+1)$ only on the right-hand-side of the equality and all the couplings are in the expression on the left-hand-side. It implies that the phase boundaries for different spins will collapse onto a single boundary surface in the space of couplings rescaled by a factor of $S(S+1)$ [for example, $\eta_x S(S+1)$ as so on]. Moreover, we can access the so-called ``classical'' limit by making the right-hand-side of the equality in Eq.~(\ref{eq:H1QPBoundary}) zero (that is, $S\rightarrow \infty$). Below we will see that even in the classical limit, one finds a finite region of phase diagram in which the {\em quantum mechanical} singlet dimer phase survives! This seems to happens when the frustration is high, or when the problem is {\em sufficiently} quasi one dimensional.
\subsection{Calculations and discussion}
In all our calculations, $J=1$ sets the unit of energy. Below we discuss two special cases of $H_{\tt I}$. In the first case, we set $\eta^\prime=0$. This is a model of coupled two-leg ladders and coupled dimerized chains interpolating between one another. The second case is for $\eta^\prime\ne 0$, but $\eta_x=\eta_y=\eta$. For $\eta=1$, it is the $J_1$-$J_2$ model.
\subsubsection{$\eta^\prime=0$}
\begin{figure}[t]
   \centering
   \includegraphics[width=7cm]{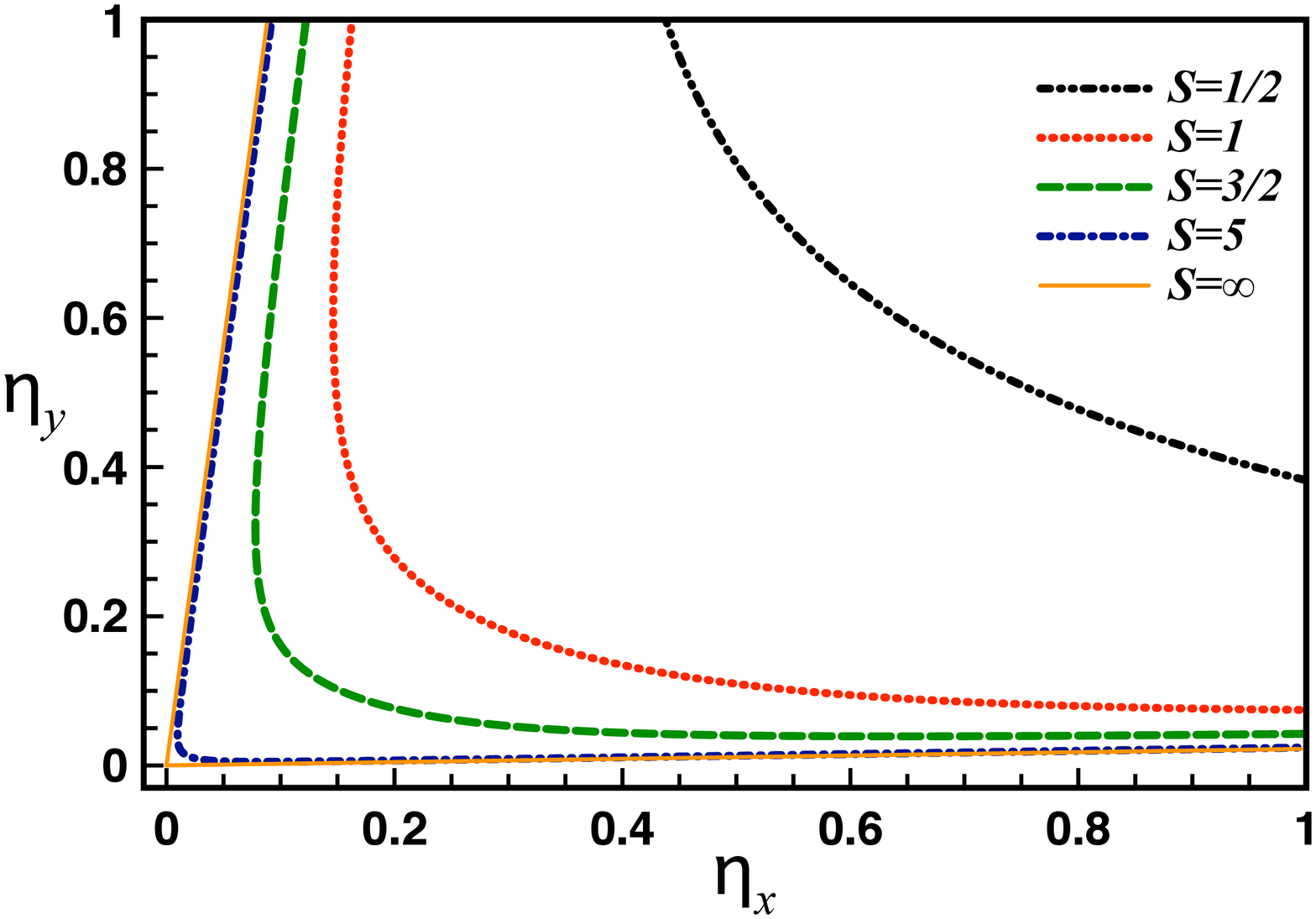}
   \includegraphics[width=7cm]{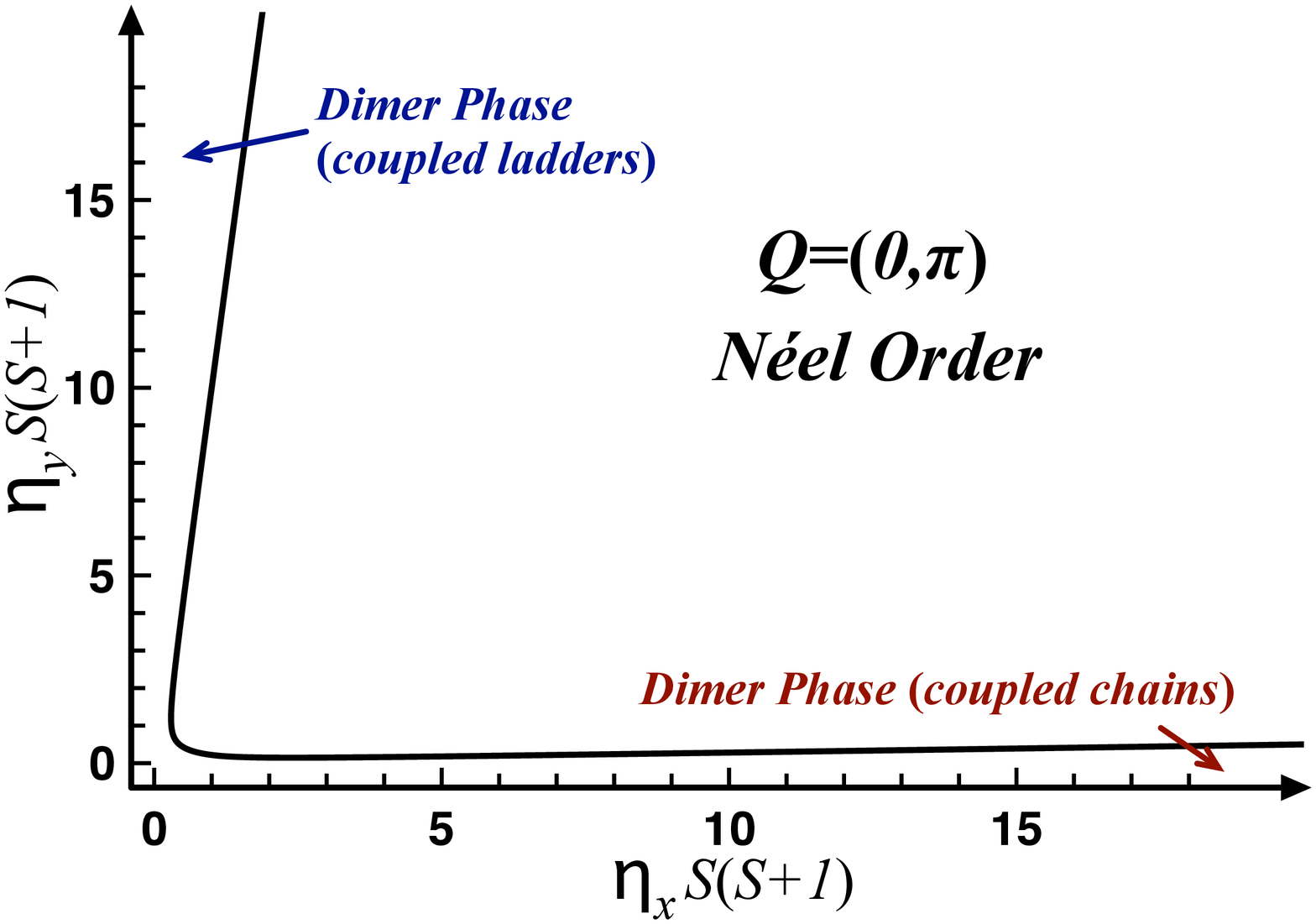}
   \caption{The mean-field quantum phase diagram of $H_{\tt I}$ for case-1 ($\eta^\prime=0$). The {\em top panel} shows the phase boundaries between the columnar-dimer and N\'eel ordered phases for different spins. For a given $S$, the region between the axes and the phase boundary is the spin-gapped dimer phase, and on the other side of the boundary is the N\'eel ordered phase. Note the quantum phase boundary in the ``classical'' limit ($S=\infty$). Surprisingly, the gapped dimer phase survives even in the classical limit for sufficiently weak $\eta_x$ (coupled two-leg ladders) or $\eta_y$ (coupled chains). In the plane of rescaled couplings ({\em bottom panel}), the quantum phase boundaries for different spins collapse onto a single line.}
   \label{fig:QPD_H1_Case1}
\end{figure}

We find $\Q=(0,\pi/a)$ in this case. Thus, it is a case of quantum phase transition from the columnar dimer to N\'eel ordered phase. We compute the phase boundaries in the $\eta_x$-$\eta_y$ plane for different values of $S$ by solving Eq.~(\ref{eq:H1QPBoundary}). Here, $\epsilon_{\Q}=\eta_x+2\eta_y$. The mean-field quantum phase diagram is presented in Fig.~\ref{fig:QPD_H1_Case1}. 
As discussed earlier, the phase boundaries for different spins collapse onto a single line in  the plane of rescaled parameters, $\eta_x\,S(S+1)$ and $\eta_y\,S(S+1)$. Fig.~\ref{fig:QPD_H1_Case1} should in principle be complemented with other calculations for a correct picture in the strongly anisotropic weak dimerization cases (for example, to have a Haldane phase for higher spins, and the like). But presently, we discuss a few interesting things about this mean-field phase diagram.
 
First about the model corresponding to $\eta_x=\eta_y=\eta$, a popular fully two-dimensional case on a dimerized square lattice. For a finite value of $S$, there occurs a quantum phase transition from the dimer to N\'eel ordered phase at some non-zero value of $\eta=\eta*$. We find that $\eta^* = \frac{0.466}{S(S+1)}$. In the limit $S\rightarrow\infty$, $\eta^*$ goes to zero however. That is, in the classical limit of this case, the ground state is N\'eel ordered even for an infinitesimally small inter-dimer coupling $\eta$.  This conforms to the usual expectations in the classical limit.  Moreover, the spin dependence of critical $\eta^*\sim \frac{1}{S(S+1)}$ is a simple analytical confirmation of a suggestion from the numerical studies of a similar model.~\cite{Richter_SpinDependence} Besides the qualitative agreement, the mean-field calculation overestimates the dimer phase in the present case. For the spin-1/2 case, $\eta^*=0.523$ from quantum Monte Carlo simulations~\cite{Matsumoto,Wenzel_Janke}, 0.535 from spin-wave analysis~\cite{Imada} and 0.54 from dimer series expansion~\cite{SinghGelfandHuse} as compared to 0.62 from the present calculation.
 
 More importantly, we want to take note of the behavior in the extremely large $S$ limit of the general case (that is, $\eta_x\ne \eta_y$). Look at the phase boundary for $S=\infty$, in the top panel of Fig.~\ref{fig:QPD_H1_Case1}, given by the lines: $\eta_y= 0.0222 \eta_x $ and  $\eta_y= 11.288 \eta_x$. Here, we find two disjoint regions (one bounded by the lines: $\eta_y=0$ and $\eta_y= 0.0222 \eta_x $, and the other by $\eta_y= 11.288 \eta_x$ and $\eta_x=0$) of the dimer phase (which is a quantum mechanical state with zero magnetic moment) existing even in this classical limit. This is a striking {\em deviation} from the normally expected behavior in the limit $S\rightarrow\infty$. The two regions can be viewed as corresponding to the quasi-1d cases of the coupled {\em dimerized} chains and the coupled two-leg ladders, respectively. It seems that the strong spatial anisotropy in the (dimerized) lattice helps the dimer state to survive even when the spins are very large. Below we will see the same behavior also recurring in the highly frustrated situations of a fully two-dimensional case (the present case of $\eta^\prime=0$ is not frustrated, but $\eta^\prime\ne 0$ in the following subsection is). While a case like $\eta_x\sim 1$ and $\eta_y\ll 1$ is known to be more sensitive towards N\'eel ordering than what the present calculation suggests ~\cite{AffleckGelfandSingh,Matsumoto,Sandvik}, we believe the existence of a dimer phase for $S\rightarrow\infty$ is very likely to come true in more accurate numerical calculations for the strongly dimerized cases ($\eta_x$, $\eta_y \ll 1$).
\subsubsection{$\eta^\prime\ne 0$ {\rm and} $\eta_x=\eta_y=\eta$}

\begin{figure}[t]
\centering
\includegraphics[width=7cm]{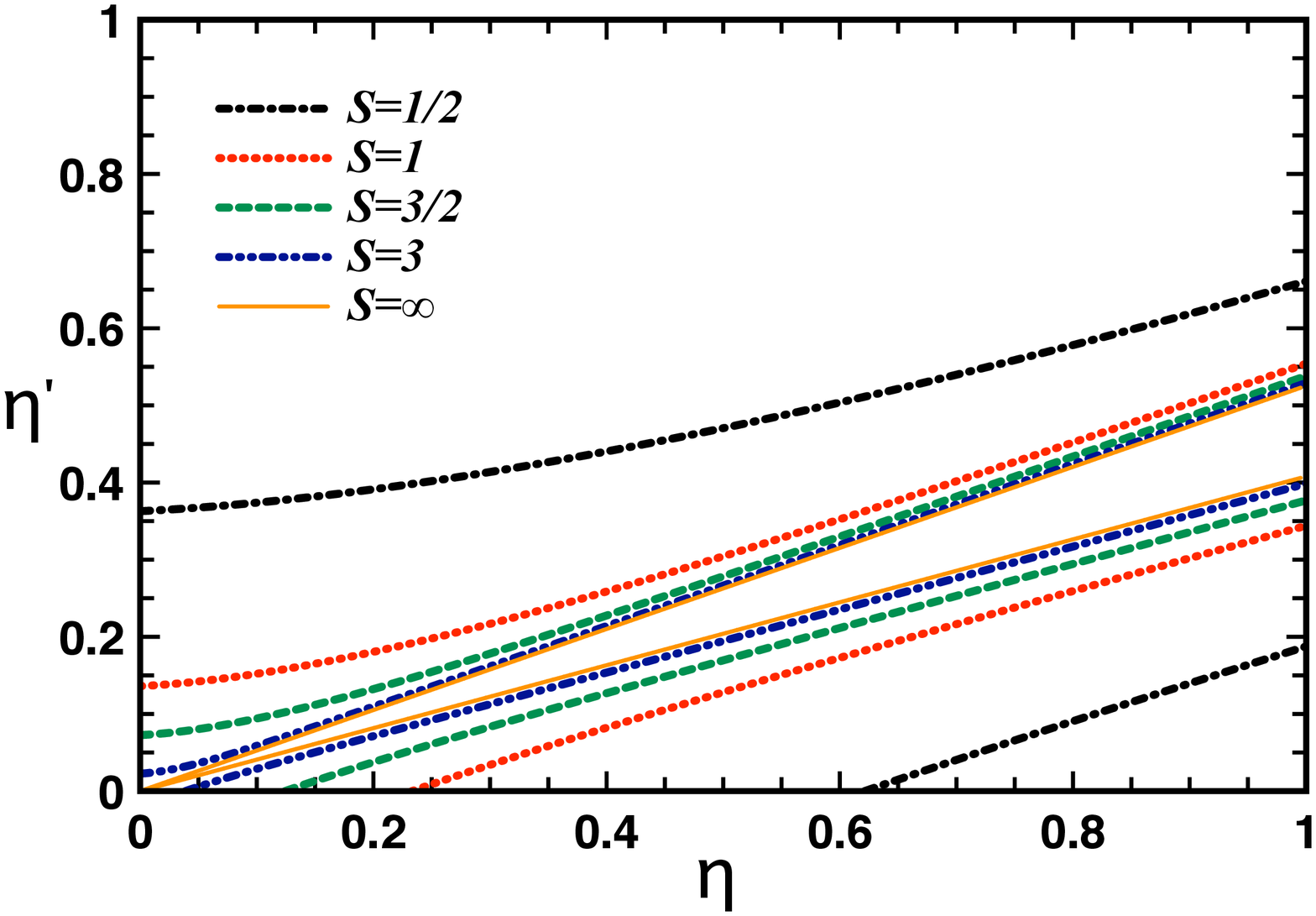}
\includegraphics[width=7cm]{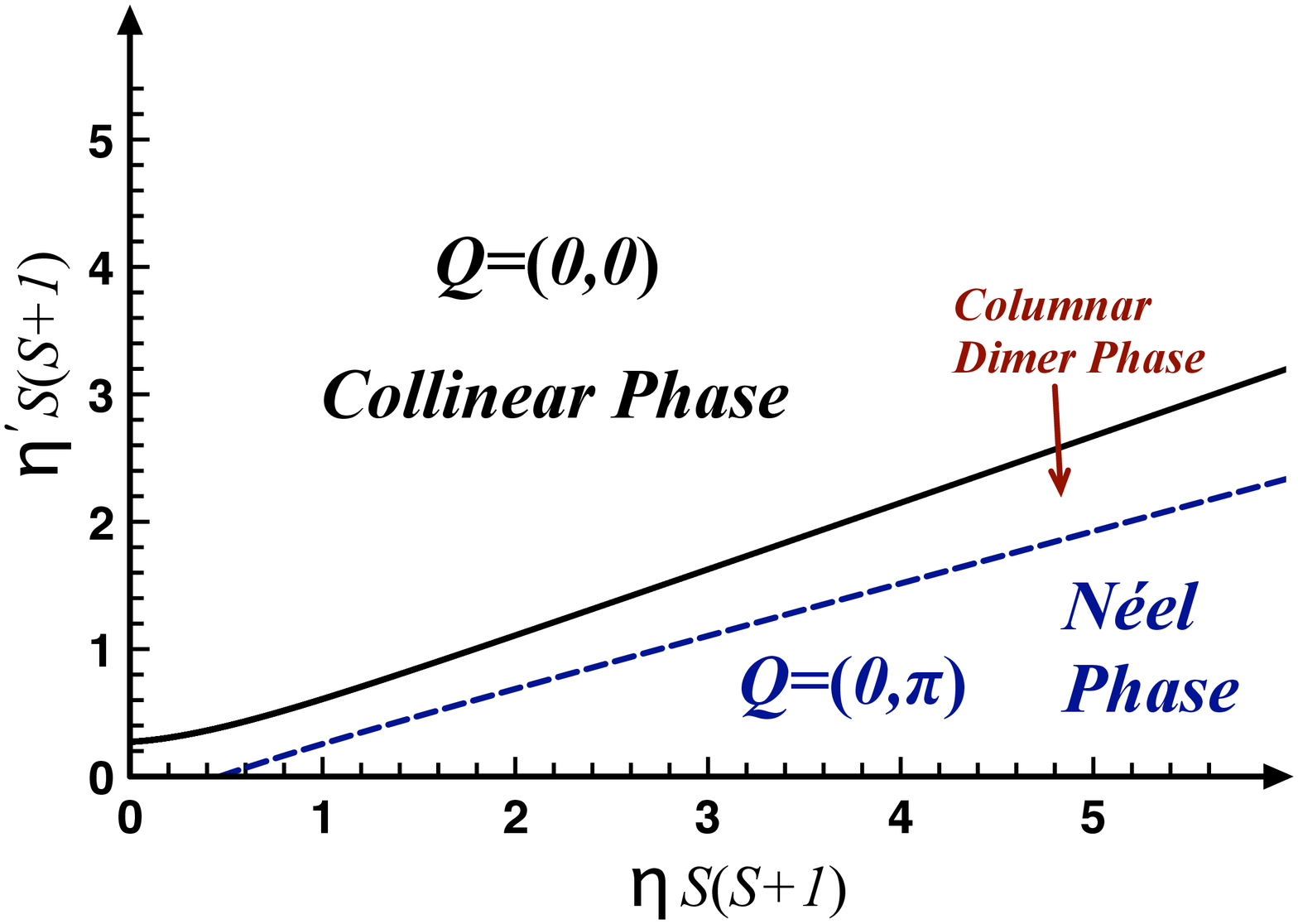}
\caption{The mean-field quantum phase diagram of $H_{\tt I}$ for case-2 ($\eta^\prime \ne 0$ and $\eta_x=\eta_y=\eta$). Here, $\eta=1$ corresponds to the $J_1$-$J_2$ model.  {\em Top panel}. For a given $S$, the region above the corresponding upper phase boundary, and bounded by the axes, is the collinear phase, while that below the lower phase boundary is the N\'eel phase. In between the two transition lines lies the columnar dimer phase.  For $S=\infty$, the phase boundaries are given by the equations, $\eta^\prime = 0.408\eta$ and $\eta^\prime=0.525\eta$.  
{\em Bottom panel}.  The phase boundaries of for different spins collapse onto two lines for two different transitions.}
\label{fig:QPD_H1_Case2}
\end{figure}

As noted above, this is the case of a frustrated two-dimensional model on a dimerized square lattice. In this case, there are two different choices of $\Q$. For weaker $\eta^\prime$, $\Q=(0,\pi/a)$, and for stronger $\eta^\prime$, it is $\Q=(0,0)$. While the former corresponds to having N\'eel order in the ground state, the latter gives collinear order (in which the magnetic moments are parallel, for the spins along the $y$-direction in Fig.~\ref{fig:model1}, and anti-parallel along the $x$-direction). From Eq.~(\ref{eq:H1QPBoundary}), we calculate the  quantum phase diagram shown in Fig.~\ref{fig:QPD_H1_Case2}. Here, $\epsilon_Q=3\eta-4\eta^\prime$ for $\Q=(0,\pi/a)$, and $4\eta^\prime-\eta$ for $\Q=(0,0)$. For $\eta^\prime=0$, the quantum critical point for different spins is given by $\eta^* = \frac{0.466}{S(S+1)}$ (same as in the previous case), and it is $\eta^{\prime *}=\frac{0.272}{S(S+1)}$ for $\eta=0$.  The phase boundaries for different spins collapse to a single line for the dimer to N\'eel transition, and similarly for the dimer to collinear transition. 

The dimer phase survives again in the limit $S\rightarrow\infty$ when the frustration is strong. To discuss this point, consider $\eta=1$ case. It corresponds to the $J_1$-$J_2$ model. In the present notation, $J_1=J$ and $J_2=\eta^\prime J$. In the classical version of this model, $\eta^\prime=0.5$ is the transition point between the N\'eel and collinear ordered ground states~\cite{j1j2_Chandra_Doucot}. It is also the point of infinitely degenerate classical ground state manifold, and hence of very high frustration. In the quantum case, for spin-1/2 specifically, it is known from many numerical studies that there exists a quantum disordered spin-gapped state (most likely a columnar dimer state) in a small range of $.4\lesssim \eta^\prime \lesssim .6$ around the $0.5$ point~\cite{vbc1,vbc2,vbc4,vbc5}. This is about $.19\lesssim \eta^\prime \lesssim .61$ from the triplon mean-field calculation~\cite{sach-bhatt,rkbk}. Below and above this range, one finds the N\'eel and collinear ordered ground state, respectively, as in the classical case.

Interestingly, even when $S$ is arbitrarily large, we find the dimer phase to be stable in a small window of $\eta^\prime$ around 0.5. For $S=\infty$ , this range is  $0.41 \lesssim \eta^\prime\lesssim 0.53$ at $\eta=1$. Away from $\eta=1$, the region of dimer phase  is bounded by the lines, $\eta^\prime = 0.408\eta$ and $\eta^\prime=0.525\eta$. Furthermore, it shrinks smoothly as one moves towards $\eta=0$. The ``quantum region'' of the phase diagram manages to survive in the classical limit seemingly because of the strong frustration. While there may be concerns about the bond-operator mean-field theory overestimating the dimer region (as it does for spin-1/2 case), a sufficient amount of frustration may always help a quantum state. Hence, we have a reasonable qualitative finding which needs to investigated further. Besides, it should be asked afresh, "is $S\rightarrow\infty$ necessarily classical?".
\section{A Model on Honeycomb lattice}
\label{sec:honeycomb}
\begin{figure}[htbp]
   \centering
   \includegraphics[width=7cm]{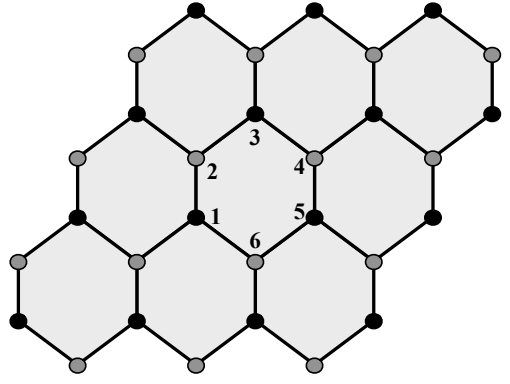}
   \caption{The model of Eq.~\ref{eq:model2}. The lines indicate the nearest neighbor Heisenberg exchange, $J$. The multiple spin-exchange proportional to $K$ is represented by a hexagon itself, with six spins labeled as 1 to 6. This multiple spin-exchange interaction is present on every hexagonal plaquette of the honeycomb. Moreover, $J$, $K>0$.}
   \label{fig:model2}
\end{figure}

We now investigate a quantum spin-$S$ model on honeycomb lattice given by the following Hamiltonian~\cite{rkdkbk}.
\begin{eqnarray}
H^{ }_{\tt II} &=&J\sum_{\langle \r, \r^\prime \rangle} \S_\r\d\S_{\r^\prime}  + \nonumber \\
&& \frac{K}{8} \sum_{\setlength{\unitlength}{0.4cm}
	\begin{picture}(1,1)	
	\put(.0,-.1){\includegraphics[width=.6cm, angle=90]{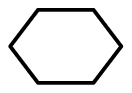}}
	\end{picture}
	} \left[\S_{12}^2\S_{34}^2\S_{56}^2 +\S_{23}^2\S_{45}^2\S_{61}^2\right] \label{eq:model2}
\end{eqnarray} 
Here, $J$ is the nearest neighbor Heisenberg interaction, and $K$ denotes the strength of a multiple-spin-exchange interaction generated by the product of pairwise total-spins of three pairs of neighboring spins on a hexagonal plaquette. The six spins on a hexagonal plaquette are labeled as 1 to 6 (see Fig.~\ref{fig:model2}). In the second term of $H_{\tt II}$, $\S_{ij}^2=(\S_i+\S_j)^2$. The interaction parameters $J$ and $K$ are taken to be positive.

An important feature of this model is that it has an exact triply degenerate dimerized singlet ground state for $J=0$ for any value of $S$. It presents an example of spontaneous dimerization in the ground state as $H_{\tt II}$ itself has no preferred dimer order (unlike $H_{\tt I}$ in the previous section). One of these dimer states corresponds to forming a singlet on every vertical nearest-neighbor bond of the honeycomb lattice in Fig.~\ref{fig:model2}. The other two states are generated from the first one by making $\pm 2\pi/3$ rotation of the lattice. For $J>0$, and for spin-1/2, we had earlier performed a triplon mean-field calculation to investigate the transition from dimer to N\'eel order in the ground state~\cite{rkdkbk}. A similar analysis was  desired for higher spins, but could not be done at that point due to the lack of spin-$S$ bond-operator representation. It was our {\em original} motivation for developing the bond-operator mean-field theory for arbitrary spins. Having achieved this objective in Sec.~\ref{sec:boS}, we can now do a triplon mean-field theory for $H_{\tt II}$, exactly in the same way as done for $H_{\tt I}$, by taking the configuration of vertical dimers on honeycomb lattice as a reference state. 

The mean-field triplon Hamiltonian in this case also looks the same as in Eq.~(\ref{eq:H1mf}). The diagnoalized mean-field Hamiltonian for $H_{\tt II}$ can therefore be written as:
\begin{eqnarray}
H_{{\tt II},mf} &=& N_d e_0 + \sum_{\k,\alpha}\sqrt{\lambda(\lambda-2\sbar^2\xi_\k)} \left(\gammahat^\dag_{\k\alpha}\gammahat^{ }_{\k\alpha} +\frac{1}{2}\right) \nonumber\\
 && + \mbox{(localized higher $j$ dimer states)},
\end{eqnarray}
where  $e_0 = -\frac{5}{2}\lambda+\sbar^2(\lambda-J-2K[S(S+1)]^2) + J[1-S(S+1)]+2K[S(S+1)]^2$, and $\xi_\k = \frac{2}{3}S(S+1)\{J+2K[S(S+1)]^2(1-\sbar^2)\}\epsilon_k$. Here, $\epsilon_\k = \cos{2\pi k_2}+\cos{2\pi(k_1-k_2)}$, and $\k$ is defined as $\k=2\pi(k_1\G_1+k_2\G_2)$ where $\G_1$ and $\G_2$ are the primitive reciprocal lattice vectors, and $ k_1, k_2 \in [0,1]$. The self-consistent equations for this problem are given by
\begin{equation}
\sbar^2 = \frac{5}{2}-\frac{3}{2N_d}\sum_\k\frac{\lambda-\sbar^2\xi_\k}{\sqrt{\lambda(\lambda-2\sbar^2\xi_\k)}}
\end{equation}
and
\begin{equation}
\lambda=J+2K[S(S+1)]^2+\frac{3}{2N_d}\sum_\k\frac{\xi_\k+\sbar^2\frac{\partial\xi_\k}{\partial\sbar^2}}{\sqrt{\lambda(\lambda-2\sbar^2\xi_\k)}}.
\end{equation}

The closing of the triplon gap, which marks the instability of the dimer phase to a magnetically ordered phase, fixes $\lambda$ as $\lambda^*=2\sbar^2\xi_\Q$, where $\Q=(0,0)$ in the present case. At this quantum critical point, we get 
\begin{subequations}
\begin{eqnarray}
\sbar^2 &=& \frac{5}{2}-\frac{3}{2N_d}\sum_k\frac{1-\frac{\epsilon_\k}{4}}{\sqrt{1-\frac{\epsilon_\k}{2}}}  \simeq 0.817,~\mbox{and} \\
\frac{J}{K} &=& 2[S(S+1)]^2\left\{-1+2\sbar^2\frac{u-\frac{4}{3}\sbar^2}{\frac{1}{S(S+1)}+u-\frac{8}{3}\sbar^2}\right\} \nonumber\\
&\simeq&2[S(S+1)]^2\left[ \frac{0.71}{1.524-\frac{1}{S(S+1)}}-1\right] \equiv \zeta^*_{S} \label{eq:JbyK}
\end{eqnarray}
\end{subequations}
where $u=\frac{1}{N_d}\sum_\k\frac{\epsilon_\k}{\sqrt{1-\frac{\epsilon_\k}{2}}} \simeq 0.655$. Equation~(\ref{eq:JbyK}) is a closed form expression for the critical $J/K$, denoted as  $\zeta^*_S$, as a function of $S$. For the dimer order to become unstable to N\'eel order in this mean-field theory, $J/K$ must be greater than $\zeta^*_S$ for a given $S$. We find $\zeta^*_{1/2} \simeq 3.067$, which agrees with our earlier calculation for the spin-1/2 case of this model. Next we find $\zeta^*_{1}\simeq -2.454$, $\zeta^*_{3/2}\simeq -12.247$, and so on. For positive $J$ and $K$, we therefore conclude that the mean-field triplon calculation predicts a N\'eel ordered ground state for $H_{\tt II}$ for $S\ge 1$ for any non-zero value of $J/K$.

We know for sure that the dimer ground state is exact for $J=0$. For the spin-1/2 case, the triplon analysis predicts that the dimer state will give way to the N\'eel state only when $J/K$ is sufficiently strong. However, for spin-1 and higher, it seems to happen for arbitrarily small $J$. At this point, it is important to note the following. While deriving the mean-field triplon Hamiltonian for $H_{\tt II}$, following the steps outlined in the previous section, we end up having {\em no contributions} from the six-spin terms of the form $(\S_1\cdot\S_2) (\S_3\cdot\S_4) (\S_5\cdot\S_6)$. For $S=1/2$, it does not seem to affect the dimer phase for small $J$ as the mean-field theory suggests. We have some evidence of this from a numerical calculation in our earlier work on this model~\cite{rkdkbk}. However, it is not clear as to how the absence of contribution from the six-spin terms in the present mean-field calculation will affect the case of spin-1 and higher.  May be, in a {\em renormalized} triplon analysis of $H_{\tt II}$, one  gets the dimer phase over a small but finite range of $J/K$ for $S\ge 1$. One such calculation is done by writing $(\S_1\cdot\S_2) (\S_3\cdot\S_4) (\S_5\cdot\S_6)$ as $-[S(S+1)]^2\sbar^2\chi \left[ \S_3\cdot\S_4+\S_5\cdot\S_6-2\chi S(S+1)\right]$, and similarly for $(\S_2\cdot\S_3) (\S_4\cdot\S_5) (\S_6\cdot\S_1)$. Here, the expectation values $\langle\S_3\cdot\S_4\rangle$, $\langle\S_5\cdot\S_6\rangle$, $\langle\S_6\cdot\S_1\rangle$ and $\langle\S_2\cdot\S_3\rangle$ are all taken to be equal to $S(S+1) \chi$. It gives the following critical value of $J/K$.
\begin{equation}
\frac{J}{K} \simeq 2[S(S+1)]^2\left[ \frac{0.856}{1.524-\frac{1}{S(S+1)}}-1\right] \equiv \zeta^*_{S} 
\end{equation}
It is similar to Eq.~(\ref{eq:JbyK}), except the numerator inside the square-brackets in different, which only slightly increases the value of $\zeta^*_{S}$. But the qualitative conclusion remains the same. That is, the N\'eel order sets in for arbitrarily small $J$ for $S\ge 1$. Well, this is the result from triplon mean-field calculation. Alternative calculations are needed to resolve this conclusively.
\section{\label{sec:extend} Extensions of the representation}
Below we present two immediate extensions of the bond-operator representation derived in Sec.~\ref{sec:boS}. First, we go beyond singlet and triplets to include quintet states on a bond. In the second case, we derive a similar representation on square plaquette in terms of the plaquette bosons, which turns out to be an easy extension of Eqs.~(\ref{eq:bo_s1}) and (\ref{eq:bo_s2}) to a plaquette problem.
\subsection{\label{sec:including_quintets} Including quintets on a bond}
The bond-operator representation derived in Sec.~\ref{sec:boS} is in the subspace of singlet and triplet states only. 
With some labor, we can extend this to include the quintets, knowing how to systematically construct the bond-eigenstates (see Appendix).  Including higher states is possible, but it requires even more effort, and will not be considered presently. The bond operator representation including quintets is written below. 
\begin{eqnarray}
S^z_{1,2} &\approx& \pm\sqrt{\frac{{\tt N}_t}{2{\tt N}_s}} \left( \shat^\dag\that^{ }_0 + \that^\dag_0 \shat \right) \pm\sqrt{\frac{{\tt N}_q}{{\tt N}_t}}\bigg[\frac{1}{\sqrt{3}}\left(\that^\dag_0\qhat^{ }_0 + \qhat^\dag_0\that^{ }_0\right) \nonumber \\
& +& \frac{1}{2}\left(\that^\dag_1\qhat^{ }_1 + \qhat^\dag_1\that^{ }_1+ \that^\dag_{\bar{1}} \qhat^{ }_{\bar{1}} + \qhat^\dag_{\bar{1}}\that^{ }_{\bar{1}}\right)\bigg] + \frac{1}{2}\Big(\that^\dag_1\that^{ }_1 - \that^\dag_{\bar{1}}\that^{ }_{\bar{1}} \nonumber \\
& +& \qhat^\dag_1\qhat^{ }_1 - \qhat^\dag_{\bar{1}}\qhat^{ }_{\bar{1}}\Big) + \left(\qhat^\dag_2\qhat^{ }_2 - \qhat^\dag_{\bar{2}}\qhat^{ }_{\bar{2}}\right) \label{eq:bo1_quintet}
\end{eqnarray}
\begin{eqnarray}
S^+_{1,2}  & \approx &  \pm\sqrt{\frac{{\tt N}_t}{{\tt N}_s}} \left( \shat^\dag\that^{ }_{\bar{1}} - \that^\dag_1 \shat \right) \pm\sqrt{\frac{{\tt N}_q}{{\tt N}_t}}\bigg[ \left(\that^\dag_{\bar{1}}\qhat^{ }_{\bar{2}} - \qhat^\dag_2\that^{ }_1\right) \nonumber \\
& & + \frac{1}{\sqrt{2}}\left(\that^\dag_0\qhat^{ }_{\bar{1}} - \qhat^\dag_1\that^{ }_0\right)+\frac{1}{\sqrt{6}}\left(\that^\dag_1 \qhat^{ }_0 - \qhat^\dag_0\that^{ }_{\bar{1}}\right)\bigg]\nonumber \\
& & + \frac{1}{\sqrt{2}}\left(\that^\dag_1\that^{ }_0 + \that^\dag_0\that^{ }_{\bar{1}}\right) + \sqrt{\frac{3}{2}}\left(\qhat^\dag_1\qhat^{ }_0 + \qhat^\dag_0\qhat^{ }_{\bar{1}}\right) \nonumber  \\
&& + \left(\qhat^\dag_2\qhat^{ }_1 + \qhat^\dag_{\bar{1}}\qhat^{ }_{\bar{2}}\right) \label{eq:bo2_quintet}
\end{eqnarray}
Here, ${\tt N}_s = 2S+1$, ${\tt N}_t= 2S(S+1){\tt N}_s/3$ and ${\tt N}_q= {\tt N}_t(2S-1)(2S+3)/5$, are the normalization constants for the singlet, triplet and quitet states respectively. Moreover, in the notation $\pm$, the `$+$' corresponds to $\S_1$ and `$-$' to $\S_2$. 

Equations~(\ref{eq:bo1_quintet}) and (\ref{eq:bo2_quintet}) are exact for spin-1 case~\cite{spin1_bo1}, and reduce to the representation for spin-1/2 operators by dropping the terms involving quintets. Note that the coefficients of the terms mixing singlet with triplets and triplets with quintets scale as $S$ for large $S$. While the strengths of different mixing terms grow similarly as $S$ grows large, the hierarchy of mixing suggests that for a spin-gapped phase in a system of exchange-interacting quatum spins, the triplon analysis is a minimal  reasonable thing to do. It is because the condensation of ``quintons'' is facilitated only by that of the triplons. In a gapped phase where triplons have not condensed, it is {\em unlikely} that the quintons will condense. Therefore, it seems okay to  ignore the quintet states to first approximation. It will not be the same however if we take into account the single anisotropy effects like $(S^z_{1,2})^2$. In this case, the singlet state will directly mix with quintets, and therefore, it will be better to work with Eqs.~(\ref{eq:bo1_quintet}) and (\ref{eq:bo2_quintet}) instead of Eqs.~(\ref{eq:bo_s1}) and (\ref{eq:bo_s2}). This combined ``tiplon-quinton'' analysis will be useful in investigating the influence of single-ion anisotropy on the stability of a dimer phase, and on its existence in the limit $S\rightarrow\infty$.
\subsection{\label{sec:plaquette} Representation on a square plaquette}
Consider a spin-$S$ problem on a single square plaquette given by the Hamiltonian: $H_{sp} = J(\S_1\cdot\S_2+\S_2\cdot\S_3+\S_3\cdot\S_4+\S_4\cdot\S_1)+J^\prime (\S_1\cdot\S_3+\S_2\cdot\S_4)$, where $J$ is the exchange interaction along the edges of the square and $J^\prime$ is the interaction along the diagonals. The subscript $sp$ stands for square plaquette. This problem can be solved by rewriting it as: 
\(H_{sp}=\frac{J}{2}\S^2_{tot} - \frac{(J-J^\prime)}{2}(\S^2_{13} +\S^2_{24})- 2J^\prime S(S+1) \), where $\S_{tot} =\S_1+\S_2+\S_3+\S_4 $ is the total spin of the plaquette, and $\S_{13}=\S_1+\S_3$ and $\S_{24}=\S_2+\S_4$ are the total spins on the two diagonals. The eigenstates of this problem are completely specified by three quantum numbers: the total spin of the plaquette, $j$, and the two diagonal spins, $j_{13}$ and $j_{24}$, with eigenvalues, $E_{sp}(j,j_{13},j_{24}) = \frac{J}{2}j(j+1) -\frac{(J-J^\prime)}{2}[j_{13}(j_{13}+1) + j_{24}(j_{24}+1)] -2J^\prime S(S+1)$. Given that we are interested in antiferromagnetic interactions ($J$, $J^\prime >0$), let us figure out the possible ground states, and derive a bosonic representation for spin operators,  considering only the lowest energy excitations.

Since $J>0$, for a given $j_{13}$ and $j_{24}$, the $E_{sp}$ would be lowest for the smallest value of $j$. Moreover, when $J>J^\prime$, the ground state of $H_{sp}$ is given by: $j_{13}=j_{24}=2S$ and $j=0$, and for $J<J^\prime$, it corresponds to $j_{13}=j_{24}=0$. The latter is a case of dimer ground state, in which the two diagonal bonds separately become singlet, and $j$ is trivially zero. The elementary excitations in this case would just correspond to making a diagonal bond a triplet. In short, for $J<J^\prime$, the bond-operator representation of Eqs.~(\ref{eq:bo_s1}) and (\ref{eq:bo_s2}) is applicable as it is. However, for $J>J^\prime$, the ground state is a true plaquette-singlet, involving all four spins. Therefore, we must separately find out a representation of the spin operators in terms of this plaquette-singlet and the corresponding plaquette-triplet excitations of the elementary kind. 

For $J>J^\prime$, the ground state lies in the sector given by $j_{13}=j_{24}=2S$. The plaquette states in this sector, for different values of $j$, are the compound eigenstates of two spins of size $2S$. That is, in Eqs.~(\ref{eq:singlet_state}) and (\ref{eq:triplet_plus1}-\ref{eq:triplet_minus1}), replace $S$ by $2S$. This immediately suggests that the diagonal spins, $\S_{13}$ and $\S_{24}$, are represented by Eqs.~(\ref{eq:bo_s1}) and (\ref{eq:bo_s2}) with $S$ written as $2S$, where $\shat$ and $\that_\alpha$ operators are now the bosons corresponding to the plaquette singlet, $|j=0;j_{13}=2S,j_{24}=2S\rangle$, and triplet states, $|j=1;j_{13}=2S,j_{24}=2S \rangle$, respectively. In order to find the representation for individual spins, $\S_1$ and $\S_3$, we must also find $\S_1-\S_3$ in terms of the plaquette bosons, and do similarly for $\S_2$ and $\S_4$. Since we consider only those states given by $j_{13}=j_{24}=2S$, the operators $\S_1-\S_3$ and $\S_2-\S_4$ would be null operators in this restricted subspace because they change the values of $j_{13}$ and $j_{24}$. Hence, their matrix elements in the subspace of the ground state singlet and the lowest triplets are zero. It leads to the following representation of the spin operators on the plaquette.
\begin{subequations}
\begin{eqnarray}
{\bf S}_{1\alpha} &=& \S_{3\alpha} \nonumber \\
&\approx& \sqrt{\frac{S(2S+1)}{6}}\left(\shat^\dag\that^{ }_\alpha + \that^\dag_\alpha\shat\right)-\frac{i}{4}\epsilon_{\alpha\beta\gamma}\that_\beta^\dag\that^{ }_\gamma \label{eq:bo_s13}\\
{\bf S}_{2\alpha} &=& \S_{4\alpha} \nonumber \\
&\approx& -\sqrt{\frac{S(2S+1)}{6}}\left(\shat^\dag\that^{ }_\alpha + \that^\dag_\alpha\shat\right)-\frac{i}{4}\epsilon_{\alpha\beta\gamma}\that_\beta^\dag\that^{ }_\gamma
\label{eq:bo_s24}
\end{eqnarray}
\end{subequations}
The above equations are written in the standard notation, except that the bosons are now defined on a square plaquette. It correctly reproduces the representation for spin-1/2 case~\cite{plaquette1}. While it is an approximate representation, it provides a simple framework for discussing the low energy physics of a (coupled) plaquette problem for $J$ sufficiently stronger than $J^\prime$ (and other couplings in a given problem). However, when $J^\prime$ is strong enough, the states from other sectors begin to compete. For example, on a single plaquette for $J^\prime > (1-\frac{1}{4S})J$, the singlet state for $j_{13}=j_{24}=2S-1$ becomes lower in energy than the triplets in the sector containing ground state. This renders the above representation insufficient for an effective low energy description. It is, in any case, a useful representation, if considered within limits.
\section{Conclusion}
\label{sec:summary}
To summarize, we have derived the bond-operator representation for spin-S dimer problems, and also worked out a similar representation on a square plaquette.  Using this bond-operator representation, we have done the mean-field triplon analysis of two model quantum antiferromagnets: 1) a coupled columnar dimers model on square lattice, and 2) a model on honeycomb lattice with spontaneous dimer order in the ground state. Through this mean-field calculation, we have studied the quantum phase transition from the dimer to AF ordered phases as a function of spin. A notable outcome of this analysis is that one finds the dimer phase, which is a quantum mechanical phase, to exist even in the limit $S\rightarrow\infty$, under the conditions of strong frustration (or spatial anisotropy with strong dimerization). It suggests that the limit $S\rightarrow\infty$ is not necessarily ``classical'', as there may not exist any ground state with non-zero classical magnetic moments for a system of (frustrated) quantum spins. Such quantum ground states are known to exist for arbitrarily large spins in specially constructed models, such as the Shastry-Sutherland model~\cite{SS} or the exactly solvable case of the model on honeycomb lattice in Sec.~\ref{sec:honeycomb}. However,  we believe this behavior of having quantum states in the so-called classical limit to occur more generically. The present observations offer an interesting view on the classical limit of frustrated quantum antiferromagnets, which further  needs to be investigated carefully.

\appendix*
\section{Compounding a pair of spins}
Here, we describe an interesting approach, that we have developed, for adding two quantum spins using the Schwinger-boson representation for spin operators. It generates the closed-form expressions for the Clebsch-Gordan coefficients rather conveniently. 

Let $\ahat$ and $\bhat$ denote the Schwinger-boson operators, in terms of which, the operators of a spin can be written as: $\S^+ = \ahat^\dag\bhat$, $\S^-=\bhat^\dag\ahat$, and $\S_z=(\ahat^\dag\ahat-\bhat^\dag\bhat)/2$, subjected to the constraint: $\ahat^\dag\ahat+\bhat^\dag\bhat=2S$. This is called the Schwinger-boson representation. Here, $\S^\pm$ and $\S_z$ are the usual spin operators, and $S$ is the spin quantum number.

For a pair of spins $\S_1$ and $\S_2$, define an antisymmetric pair-operator (also called the valence-bond operator)~\cite{fnote_AKLT}:
\begin{equation} 
\calA^\dag= \ahat^\dag_1\bhat^\dag_2-\bhat^\dag_1\ahat^\dag_2, \label{eq:calAdag}
\end{equation}
and three symmetric pair-operators:
\begin{subequations}
\begin{eqnarray}
 \calB_\one^\dag &=& \ahat^\dag_1\ahat^\dag_2 \label{eq:calB1dag}\\
 \calB_\zero^\dag &=& \ahat^\dag_1\bhat^\dag_2+\bhat^\dag_1\ahat^\dag_2 \label{eq:calB0dag}\\
 \calB_\onebar^\dag &=& \bhat^\dag_1\bhat^\dag_2 \label{eq:calB1bardag}
\end{eqnarray}
\end{subequations}
where the subscripts $1$ and $2$ stand for the two spins (or, two sites of a bond). These pair-operators will form the basis of our analysis for evaluating the total-spin eigenstates of a pair of spins. Besides, we will also use the following operator identities for systematic proofs.
\begin{equation}
[\calA,\calA^\dag] = 2+\calN_1+\calN_2
\label{eq:commuteAAdag}
\end{equation}
Here, $\calN_1=\ahat^\dag_1\ahat^{ }_1+\bhat^\dag_1\bhat^{ }_1$, and $\calN_2=\ahat^\dag_2\ahat^{ }_2+\bhat^\dag_2\bhat^{ }_2$ are the number operators of the Schwinger bosons on site 1 and 2, respectively. By successive application of the above relation, we can show that
\begin{eqnarray}
\calA^\dag \calA \left(\calA^\dag\right)^l |n_1,n_2\rangle &=&\left(\calA^\dag\right)^l \Big[l(n_1+n_2+l+1) \nonumber \\ 
&&~~~~~~~~ + \calA^\dag\calA \Big] |n_1,n_2\rangle \label{eq:AdagAAdagl}
\end{eqnarray}
where $|n_1,n_2\rangle$ denotes a state with total number of $n_1$ Schwinger bosons on site 1, and $n_2$ on site 2. The other useful relations are:
\begin{equation}
2\S_1\cdot\S_2 = \frac{1}{2}\calN_1\calN_2-\calA^\dag\calA
\label{eq:S1dotS2}
\end{equation}
and
\begin{equation}
\left(\S_1+\S_2\right)^2 = \frac{1}{4}(\calN_1+\calN_2)(\calN_1+\calN_2+2)-\calA^\dag\calA.
\label{eq:S1plusS2square}
\end{equation}
We will also need the following commutators.
\begin{subequations}
\begin{eqnarray}
\left[\calA,(\ahat^\dag_1)^l \right] &=& l (\ahat^\dag_1)^{l-1} \bhat_2 \label{eq:commuteAa1dagl}\\
\left[\calA,(\bhat^\dag_1)^l \right] &=& -l (\bhat^\dag_1)^{l-1} \ahat_2 \label{eq:commuteAb1dagl}\\
\left[\calA,(\ahat^\dag_2)^l \right] &=& -l (\ahat^\dag_2)^{l-1} \bhat_1 \label{eq:commuteAa2dagl}\\
\left[\calA,(\bhat^\dag_2)^l \right] &=& l (\bhat^\dag_2)^{l-1} \ahat_1 \label{eq:commuteAb2dagl}
\end{eqnarray}
\end{subequations}
Now we are all set to construct the total-spin eigenstates, $|j,m_j\rangle$, where $j$ denotes the total-spin quantum number, and $m_j$ is the quantum number for the $z$-component of the total spin for a given $j$.

\subsection{\label{subsec:equal_spin} Case of equal spins: $S_1=S_2=S$}
$\bullet$ {\bf\em Proposition 1.} The normalized singlet eigenstate is given by: \[ |j=0,m_j=0\rangle = \frac{1}{(2S)!\sqrt{2S+1}}\left(\calA^\dag\right)^{2S}|0,0\rangle. \]

{\em Proof.}  It is clear that $n_1=n_2=2S$ in the proposed state, $\left(\calA^\dag\right)^{2S}|0,0\rangle$, where $|0,0\rangle$ denotes the Schwinger-boson vacuum in which $n_1=n_2=0$.

Apply $(\S_1+\S_2)^2$ on the proposed state. Using Eqs.~(\ref{eq:S1plusS2square}) and (\ref{eq:AdagAAdagl}), we find that \[ (\S_1+\S_2)^2\left(\calA^\dag\right)^{2S}|0,0\rangle =0. \]  Since the proposed state is annihilated by the total-spin operator, it is a singlet. That is, $j=0$. Moreover, $(\S_1+\S_2)_z \left(\calA^\dag\right)^{2S}|0,0\rangle =0$, because $\calA^\dag$ changes the total number of $a$-type and $b$-type Schwinger bosons by the same amount (that is, one). Therefore, $m_j=0$. Now, we fix the normalization.

Let, $\mbox{\tt Norm}[2S]=\langle 0,0| (\calA)^{2S}\left(\calA^\dag\right)^{2S}|0,0\rangle$, be the normalization constant. Clearly, $\mbox{\tt Norm}[0]=1$. Moreover, we find that \[\mbox{\tt Norm}[2S]=(2S+1).2S.\mbox{\tt Norm}[2S-1].\] It is derived using $\calA\calA^\dag=(2+\calN_1+\calN_2)(4+\calN_1+\calN_2)/4 - (\S_1+\S_2)^2$. This recursive relation for normalization implies, $\mbox{\tt Norm}[2S]=(2S+1)[(2S)!]^2$. Hence, the proof. $\bullet$

Next we workout a procedure for generating the eigenstates for arbitrary $j$. Let us introduce a ``generating'' operator, $\calB^\dag(\xi)=\calB^\dag_\one + \xi\calB^\dag_\zero +\xi^2\calB^\dag_\onebar$, where $\xi$ is just a parameter. It can also be written as: $\calB^\dag(\xi)=(\ahat^\dag_1+\xi\bhat^\dag_1)(\ahat^\dag_2+\xi\bhat^\dag_2)$.

$\bullet$ {\bf\em Proposition 2.} The total-spin quantum number of the state, \(|j;\xi \rangle=\left[\calB^\dag(\xi)\right]^j\left(\calA^\dag\right)^{2S-j}|0,0\rangle\), is $j$.

{\em Proof.} Evaluate $ (\S_1+\S_2)^2|j;\xi\rangle $ as follows.\\
\begin{eqnarray}
&&(\S_1+\S_2)^2|j;\xi\rangle \nonumber\\
 &&=  \left[2S(2S+1)-\calA^\dag\calA\right]\left(\calA^\dag\right)^{2S-j}\left[\calB^\dag(\xi)\right]^j|0,0\rangle \nonumber \\
&& \hspace{1cm} {\mbox{[Equation~(\ref{eq:AdagAAdagl}) implies the following]}} \nonumber\\
 &&= j(j+1)|j;\xi\rangle-\calA^{2S-j+1}\calA\left[\calB^\dag(\xi)\right]^j|0,0\rangle \nonumber
\end{eqnarray}
For $|j;\xi\rangle$ to be an eigenstate of the total-spin operator with eigenvalue $j(j+1)$, the operator $\calA$ must annihilate $\left[\calB^\dag(\xi)\right]^j|0,0\rangle$. Below, we show that it is true.
\begin{eqnarray}
&& \calA\left[\calB^\dag(\xi)\right]^j|0,0\rangle =\nonumber\\
&&\sum_{l_1, l_2=0}^j C^j_{l_1} C^j_{l_2}\xi^{l_1+l_2}\calA\,(\ahat^\dag_1)^{j-l_1}(\bhat^\dag_1)^{l_1}(\ahat^\dag_2)^{j-l_2}(\bhat^\dag_2)^{l_2} |0,0\rangle \nonumber \\
&& \hspace{1cm}{\mbox{[Eqs.~(\ref{eq:commuteAa1dagl}) \& (\ref{eq:commuteAb1dagl}) imply the following.]}} \nonumber \\
&& =j^2\xi[\calB^\dag(\xi)]^{j-1}|0,0\rangle + \nonumber\\
&&\sum_{l_1, l_2=0}^j C^j_{l_1} C^j_{l_2}\xi^{l_1+l_2}(\ahat^\dag_1)^{j-l_1} \calA\,(\bhat^\dag_1)^{l_1}(\ahat^\dag_2)^{j-l_2}(\bhat^\dag_2)^{l_2} |0,0\rangle \nonumber\\
&& =\underbrace{j^2\xi[\calB^\dag(\xi)]^{j-1}|0,0\rangle -j^2\xi[\calB^\dag(\xi)]^{j-1}|0,0\rangle}+\nonumber\\
&&\sum_{l_1, l_2=0}^j C^j_{l_1} C^j_{l_2}\xi^{l_1+l_2}(\ahat^\dag_1)^{j-l_1}(\bhat^\dag_1)^{l_1}\underbrace{\calA(\ahat^\dag_2)^{j-l_2}(\bhat^\dag_2)^{l_2} |0,0\rangle} \nonumber \\
&& =0 \nonumber
\end{eqnarray}
Here, the quantities above the under-braces are zero (first one due to a simple cancellation, and second because the position of $\calA$ is such that $\ahat_1$ and $\bhat_1$ annihilate $|0,0\rangle$). The relation, $\calA\left[\calB^\dag(\xi)\right]^j |0,0\rangle=0$, straightforwardly implies \((\S_1+\S_2)^2|j;\xi\rangle= j(j+1)|j;\xi\rangle\).
Hence, the proof. $\bullet$

Below we show that all the different $m_j$ states for a given $j$ are contained in $|j;\xi\rangle$, and express themselves in powers of $\xi$. Since all $|j,m_j\rangle$ states can be systematically derived from $|j;\xi\rangle$, we call $|j;\xi\rangle$ the ``generating state''. Note that $|j;\xi\rangle$ is not a normalized state.

$\bullet$ {\bf\em Proposition 3.} The generating state, $|j;\xi\rangle$, has the following series expansion in powers of $\xi$. \[|j;\xi\rangle=\sum_{m_j=j}^{-j}\xi^{j-m_j}|\widetilde{j,m_j}\rangle\]
Here, $|\widetilde{j,m_j}\rangle$ denotes an unnormalized total-spin eigenstate (as compared to $|j,m_j\rangle$, which denotes the normalized version of $|\widetilde{j,m_j}\rangle$).

{\em Proof.} To derive the series form of $|j;\xi\rangle$, expand $\left[\calB^\dag(\xi)\right]^j$ in powers of $\xi$.
\begin{eqnarray}
&& \left[\calB^\dag(\xi)\right]^j = \left(\calB^\dag_\one + \xi\calB^\dag_\zero +\xi^2\calB^\dag_\onebar\right)^j \nonumber \\
&& =\sum_{l_1=0}^j\sum_{l_2=0}^{l_1} C^j_{l_1}C^{l_1}_{l_2} \xi^{l_1+l_2} (\calB^\dag_\one)^{j-l_1}(\calB^\dag_\zero)^{l_1-l_2}(\calB^\dag_\onebar)^{l_2} \nonumber
\end{eqnarray}
Note that $\calB^\dag_\one$ contributes $1$ to $(\S_1+\S_2)_z$, $\calB^\dag_\zero$ contributes zero, and $\calB^\dag_\onebar$ contributes $-1$ to the same. Moreover, $\calA^\dag$ adds nothing to it. Therefore, the eigenvalue of $(\S_1+\S_2)_z$ for the terms corresponding to $\xi^{l_1+l_2}$ in the generating state, $|j;\xi\rangle$, is equal to $j-l_1-l_2$. Make a transformation of the summation variables from $(l_1,l_2)$ to $(m_j,l)$ such that $m_j=j-l_1-l_2$ and $l=l_2$. Or conversely, $l_1=j-m_j-l$, and $l_2=l$, where $m_j = j\rightarrow -j$ and $l = \mbox{max}(0,-m_j)\rightarrow \left[\frac{j-m_j}{2}\right]$. Here, $[x]$ denotes the integer-valued part of $x$. Now, we can write:
\[ \left[\calB^\dag(\xi)\right]^j=\sum_{m_j=j}^{-j}\xi^{j-m_j}\calB^\dag(j,m_j) \] 
where
\begin{eqnarray}
\calB^\dag(j,m_j) &=& \sum_{l=\mbox{max}(0,-m_j)}^{\left[\frac{j-m_j}{2}\right]} \frac{j!}{(m_j+l)! (j-m_j-2l)! \,l!} \nonumber \\ 
&& ~~ \times (\calB^\dag_\one)^{m_j+l} \, (\calB^\dag_\zero)^{j-m_j-2l} \, (\calB^\dag_\onebar)^{l} \label{eq:Bdag_jmj}
\end{eqnarray}

The (unnormalized) generating state is thus written as: $|j;\xi\rangle=\sum_{m_j=j}^{-j}\xi^{j-m_j}|\widetilde{j,m_j}\rangle$, where 
\[|\widetilde{j,m_j}\rangle = \calB^\dag(j,m_j)\left(\calA^\dag\right)^{2S-j}|0,0\rangle.\]

It has been argued above that $(\S_1+\S_2)_z|\widetilde{j,m_j}\rangle = m_j|\widetilde{j,m_j}\rangle$. Therefore, $\langle\widetilde{j,m_j}|\widetilde{j,m^\prime_j}\rangle=0$ for $m_j\neq m^\prime_j$. Furthermore, Proposition 1 implies \[\sum_{m_j=j}^{-j}\xi^{j-m_j}\left[(\S_1+\S_2)^2-j(j+1)\right]|\widetilde{j,m_j}\rangle =0\] which in turn implies $(\S_1+\S_2)^2|\widetilde{j,m_j}\rangle=j(j+1)|\widetilde{j,m_j}\rangle$. Hence the proof. $\bullet$

The above mathematical result can be understood in the following way. Think of the singlet state with $2S$ valence bonds as a reference state, kind of a ``valence-bond sea''. Out of which, one can generate different $|j,m_j\rangle$ states by removing $j$ valence bonds, and inserting the same number of symmetrized bonds in a suitable way. This insertion is precisely given by the operator $\calB^\dag(j,m_j)$ defined in Eq.~(\ref{eq:Bdag_jmj}). For example, the triplet eigenstates ($j=1$ and $m_j=1,0,-1$) can be constructed as: 
\begin{subequations}
\begin{eqnarray}
&& |1,1\rangle \propto \calB^\dag_\one \left(\calA^\dag \right)^{2S-1}|0,0\rangle \label{eq:1plus1}\\
&& |1,0\rangle \propto \calB^\dag_\zero \left( \calA^\dag \right)^{2S-1}|0,0\rangle \label{eq:1zero}\\
&& |1,-1\rangle \propto \calB^\dag_\onebar \left( \calA^\dag \right)^{2S-1}|0,0\rangle \label{eq:1minus1}
\end{eqnarray}
\end{subequations}
\begin{figure}[htbp]
\centering
\includegraphics[width=6.75cm]{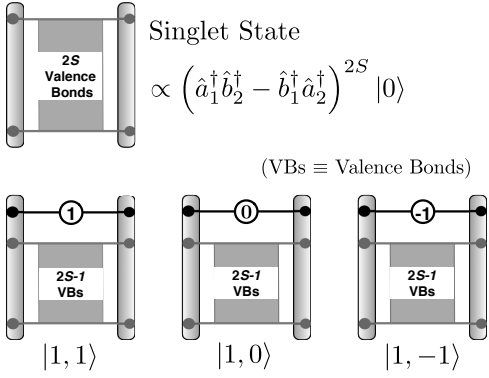}
\caption{(Top) The singlet state for a pair of spin-$S$. It can be viewed as a ``sea'' of $2S$ valence bonds. (Bottom) The three triplet states are created by replacing one valence-bond by three different symmetric bonds. See the text for details. }
\label{fig:singlet_triplets}
\end{figure}
This procedure is  pictorially illustrated in Fig.~\ref{fig:singlet_triplets}. 

\begin{figure}[htbp]
   \centering
   \includegraphics[width=6.25cm]{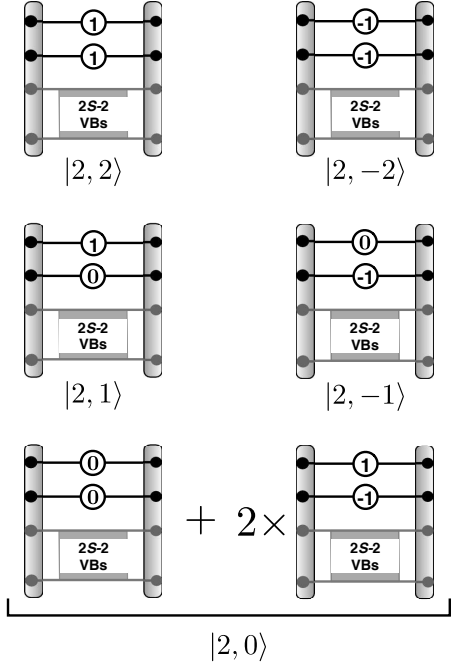}
   \caption{The quintet states. See Eqs.~(\ref{eq:2plus2}) to (\ref{eq:2zero}).}
   \label{fig:quintets}
\end{figure}

Similarly, the eigenstates for $j=2$ and $m_j=2,1,0$ are given by (see Fig.~\ref{fig:quintets}):
\begin{subequations}
\begin{eqnarray}
 |2,2\rangle &\propto& \left(\calB^\dag_\one\right)^2 \left(\calA^\dag \right)^{2S-2}|0,0\rangle \label{eq:2plus2}\\
  |2,1\rangle &\propto& \calB^\dag_\one \calB^\dag_\zero \left( \calA^\dag \right)^{2S-2}|0,0\rangle \label{eq:2plus1}\\
 |2,0\rangle &\propto& \left[ (\calB^\dag_\zero)^2+2\calB^\dag_\one \calB^\dag_\onebar\right] \left( \calA^\dag \right)^{2S-2}|0,0\rangle \label{eq:2zero}
\end{eqnarray}
\end{subequations}
The states for $m_j=-1$ and $-2$ can be obtained by replacing $\calB^\dag_\one$ by $\calB^\dag_\onebar$ in the equations for $|2,1\rangle$ and $|2,2\rangle$, respectively.

While Eq.~(\ref{eq:Bdag_jmj}) can be nicely visualized, and is helpful in understanding the generating procedure, it is not the most convenient form of $\calB^\dag(j,m_j)$. Below we present a more effective form of this operator for evaluating $|j,m_j\rangle$.
\begin{eqnarray}
&& \calB^\dag(j,m_j) = \sum_{l=\mbox{max}(0,-m_j)}^{\mbox{min}(j,j-m_j)}  C^{j}_{m_j+l}C^{j}_l  \nonumber \\ 
&&  ~~\times (\ahat^\dag_1)^{m_j+l} \, (\bhat^\dag_1)^{j-m_j-l} \, (\ahat^\dag_2)^{j-l} \, (\bhat^\dag_2)^l \label{eq:Bdag_jmj_2ndform}
\end{eqnarray}
The above expression is derived by noting that $\calB^\dag(\xi)=(\ahat^\dag_1+\xi\bhat^\dag_1)(\ahat^\dag_2+\xi\bhat^\dag_2)$. Having learnt the generating state description of the total-spin eigenstates, we now find the normalization constant, $\langle\widetilde{j,m_j}|\widetilde{j,m_j}\rangle$.

$\bullet$ {\bf\em Proposition 4.} The normalized total-spin eigenstate, with quantum numbers $j$ and $m_j$, is given by: \[ |j,m_j\rangle=\frac{1}{j! \, (2S-j)!\sqrt{C^{2S+j+1}_{2j+1} C^{2j}_{j-m_j}}} |\widetilde{j,m_j}\rangle \] where $|\widetilde{j,m_j}\rangle$ are the corresponding unnormalized states in {\em Proposition 3}.

{\em Proof.} Following the same steps as for the normalization of the singlet state in {\em Proposition 1}, we find that 
\begin{eqnarray}
&& \langle j;\xi| j;\xi\rangle \nonumber \\
&& = C^{2S+j+1}_{2j+1} \, \left[ (2S-j)! \right]^2 \langle 0,0|\left[\calB(\xi)\right]^j \left[\calB^\dag(\xi)\right]^j |0,0 \rangle \nonumber
\end{eqnarray}
We further find that
\[ \langle 0,0|\left[\calB(\xi)\right]^j \left[\calB^\dag(\xi)\right]^j |0,0\rangle = \left[j! (1+\xi^2)^j\right]^2. \] Therefore,  
\[ \langle j;\xi| j;\xi\rangle = C^{2S+j+1}_{2j+1} \, \left[ j! \, (2S-j)! \right]^2 \left(1+\xi^2\right)^{2j}. \]
Moreover, $ \langle j;\xi| j;\xi\rangle=\sum_{m_j=j}^{-j} \xi^{2(j-m_j)}\langle\widetilde{j,m_j}|\widetilde{j,m_j}\rangle$ (as implied by {\em Proposition 3}). Therefore, 
\[ \langle\widetilde{j,m_j}|\widetilde{j,m_j}\rangle = C^{2S+j+1}_{2j+1}C^{2j}_{j-m_j} \, \left[ j! \, (2S-j)! \right]^2 \]
and the normalized total-spin eigenstate, $|j,m_j\rangle$ is:
\[ |j,m_j\rangle=\frac{1}{j! \, (2S-j)!\sqrt{C^{2S+j+1}_{2j+1} C^{2j}_{j-m_j}}} |\widetilde{j,m_j}\rangle.\]
Hence, the proof. $\bullet$

Finally, we derive the explicit expression for $|j,m_j\rangle$ in terms of the actual spin states. It will give us all the Clebsh-Gordan coefficients in a closed form for artbirary $j$ and $m_j$ for a given pair of spin-$S$.

$\bullet$ {\bf\em Proposition 5.} The total-spin eigenstates, $|j,m_j\rangle$, can be written as:
\begin{eqnarray}
&&|j,m_j\rangle = \sum_{m=0}^{2S-|m_j|} \mathbb{C}[j,m_j;m] \nonumber \\
&& \hspace{5mm} \times \left\{
                         \begin{array}{ll} |S-m,-S+m+m_j\rangle &~ \forall ~  0\le m_j\le j\\ &\\
                                                      |S-m+m_j,-S+m\rangle &~ \forall ~ -j\le m_j \le 0
                          \end{array} \right. \nonumber
\end{eqnarray}
where the {\em Clebsch-Gordan} coefficients, $\mathbb{C}[j,m_j;m]$, are given by~\cite{fnote_CG}:
\begin{eqnarray}
&& \mathbb{C}[j,m_j;m] = \frac{C^{2S}_j}{\sqrt{ C^{2S+j+1}_{2j+1} C^{2j}_{j+|m_j|}  C^{2S}_{m+|m_j|}  C^{2S}_m }} \nonumber \\
&& \times \sum_{p=\mbox{max}(0,m+|m_j|-j)}^{\mbox{min}(2S-j,m)}(-)^p C^{2S-j}_p C^j_{m-p} C^j_{m+|m_j|-p} \nonumber
\end{eqnarray}
The states, $|S-m,-S+m+m_j\rangle$, above denote the product states, $|S1=S, m_1=S-m\rangle \otimes |S_2=S,m_2=-S+m+m_j\rangle$,  of the two spins.

{\em Proof.} Consider $\calB^\dag(j,m_j)$ as given in Eq.~(\ref{eq:Bdag_jmj_2ndform}). We discuss the positive and negative values of $m_j$ separately.

For $0\le m_j\le j$, 
\begin{eqnarray}
&& \calB^\dag(j,m_j) \nonumber  \\
&& =\sum_{l=0}^{j-m_j} C^{j}_{m_j+l}C^{j}_l (\ahat^\dag_1)^{m_j+l} (\bhat^\dag_1)^{j-m_j-l} (\ahat^\dag_2)^{j-l} (\bhat^\dag_2)^l \nonumber
\end{eqnarray}
Therefore,
\begin{eqnarray}
&& |\widetilde{j,m_j}\rangle = \calB^\dag(j,m_j)\left(\calA^\dag\right)^{2S-j}|0,0\rangle \nonumber \\
&&=\sum_{l=0}^{j-m_j}\sum_{p=0}^{2S-j}(-)^p C^{j}_{m_j+l}C^{j}_{l} C^{2S-j}_{p} (\ahat^\dag_1)^{(m_j+l+2S-j-p)} \nonumber \\
&& \times (\bhat^\dag_1)^{(j-m_j-l+p)} \, (\ahat^\dag_2)^{(j-l+p)} \, (\bhat^\dag_2)^{(l+2S-j-p)} |0,0\rangle \nonumber \\
&& =(2S)!\sum_{l=0}^{j-m_j}\sum_{p=0}^{2S-j}(-)^p \frac{C^{j}_{m_j+l}C^{j}_{l} C^{2S-j}_{p} }{\sqrt{C^{2S}_{j-m_j-l+p}C^{2S}_{j-l+p}}} \nonumber \\
&& \hspace{1.5cm} \times |S-j-p+l+m_j,-S+j+p-l\rangle \nonumber \\
&& [\mbox{change of variable:} ~ l\rightarrow j-m_j-l] \nonumber\\
&& =(2S)!\sum_{l=0}^{j-m_j}\sum_{p=0}^{2S-j}(-)^p \frac{C^{j}_{l} C^{j}_{l+m_j} C^{2S-j}_{p} }{\sqrt{C^{2S}_{l+p} C^{2S}_{l+p+m_j}}} \nonumber \\
&& \hspace{1.5cm} \times |S-l-p,-S+l+p+m_j\rangle \nonumber
\end{eqnarray}
Define new variables $m$ and $\bar{p}$ as: $m=l+p,~\bar{p}=p$. Now, we can write the normalized state, $|j,m_j\rangle$, as:
\begin{eqnarray}
&& |j,m_j\rangle = \frac{C^{2S}_{j}}{\sqrt{C^{2S+2j+1}_{2j+1} C^{2j}_{j+m_j}}}\sum_{m=0}^{2S-m_j} \frac{1}{\sqrt{C^{2S}_{m} C^{2S}_{m+m_j}}} \nonumber \\
&&\times \left(\sum_{\bar{p}=\mbox{max}(0,m+m_j-j)}^{\mbox{min}(2S-j,m)}(-)^{\bar{p}} C^{j}_{m-\bar{p}} C^{j}_{m+m_j-\bar{p}} C^{2S-j}_{\bar{p}} \right) \nonumber\\
&&\times  |S-m,-S+m+m_j\rangle \nonumber\\
&& = \sum_{m=0}^{2S-m_j}\mathbb{C}^{+}[j,m_j;m] |S-m,-S+m+m_j\rangle \label{eq:positive_mj_state}
\end{eqnarray}
where the coefficients of linear combination, famously called the Clebsch-Gordan coefficients, are given by
\begin{eqnarray}
&& \mathbb{C}^{+}[j,m_j;m]=\frac{C^{2S}_j}{\sqrt{C^{2S+2j+1}_{2j+1} C^{2j}_{j+m_j} C^{2S}_{m+m_j} C^{2S}_{m} }} \times  ~~~~ \nonumber \\
&& \sum_{p=\mbox{max}(0,m+m_j-j)}^{\mbox{min}(2S-j,m)}(-)^p C^{j}_{m-p} C^{j}_{m_j+m-p} C^{2S-j}_{p} 
\end{eqnarray}
Here, the superscript, $+$, indicates that it is for positive values of $m_j$. Also note the summation variable $\bar{p}$ written as $p$ (it is allowed for dummy variables). We can carry out the same analysis for the negative values of $m_j$. However, we will infer the negative $m_j$ states using an interesting argument described below.

For $-j\le m_j\le 0$, write $m_j=-|m_j|$. Then,
\begin{eqnarray}
&& \calB^\dag(j,m_j) =  \calB^\dag(j,-|m_j|)  \nonumber  \\
&& =\sum_{l=|m_j|}^{j} C^{j}_{m_j+l}C^{j}_l (\ahat^\dag_1)^{m_j+l} (\bhat^\dag_1)^{j-m_j-l} (\ahat^\dag_2)^{j-l} (\bhat^\dag_2)^l \nonumber \\
&& [\mbox{change of variable:}~ l\rightarrow l+|m_j|] \nonumber\\
&& = \sum_{l=0}^{j-|m_j|} C^{j}_{l}C^{j}_{|m_j|+l} (\ahat^\dag_1)^{l} (\bhat^\dag_1)^{j-l} (\ahat^\dag_2)^{j-|m_j|-l} (\bhat^\dag_2)^{|m_j|+l} \nonumber
\end{eqnarray}
Comparing the last line of the above equation with $\calB^\dag(j,|m^{ }_j|)$ suggests that $m_j\rightarrow -m_j$ is equivalent to the mapping: $\ahat_1\leftrightarrow \bhat_2$ and $\bhat_1\leftrightarrow\ahat_2$. Under this mapping, the valence-bond operator $\calA$ is invariant. However, the spin quantum numbers exchange: $S_1\leftrightarrow S_2$. Moreover, $\S_{1z} \leftrightarrow -\S_{2z}$. Therefore, the total-spin eigenstate for a negative $m_j$ is given by
\begin{eqnarray}
&& |j,m_j\rangle=|j,-|m_j|\rangle \nonumber \\
&& = \sum_{m=0}^{2S-|m_j|}\mathbb{C}^{+}[j,|m_j|;m] |S-m-|m_j|,-S+m\rangle ~~~~~~ \nonumber \\
&& =  \sum_{m=0}^{2S-|m_j|}\mathbb{C}^{-}[j,m_j;m] |S-m+m_j,-S+m\rangle \label{eq:negative_mj_state}
\end{eqnarray}
where the coefficient $\mathbb{C}^{-}[j,m_j;m]$, for negative $m_j$, is equal to $ \mathbb{C}^{+}[j,|m_j|;m]$.
Equations~(\ref{eq:positive_mj_state}) and (\ref{eq:negative_mj_state}) together can be stated as follows.
\begin{eqnarray}
&& |j,m_j\rangle = \sum_{m=0}^{2S-|m_j|} \mathbb{C}[j,m_j;m] \nonumber \\
&& \hspace{5mm} \times \left\{
                         \begin{array}{ll} |S-m,-S+m+m_j\rangle &~ \forall ~  0\le m_j\le j\\ &\\
                                                      |S-m+m_j,-S+m\rangle &~ \forall ~ -j\le m_j \le 0
                          \end{array} \right. \nonumber
\end{eqnarray}
Or even more compactly, 
\begin{eqnarray}
 |j,m_j\rangle = \sum_{m=0}^{2S-|m_j|} \mathbb{C}[j,m_j;m] \times \hspace{2.4cm}&& \nonumber \\
|S-m+\mbox{min}(0,m_j),-S+m+\mbox{max}(0,m_j)\rangle && \label{eq:jmj_finalform}
\end{eqnarray}
where $\mathbb{C}[j,m_j;m]= \mathbb{C}^{+}[j,|m_j|;m]$. Hence, the proof. $\bullet$

With some care, we can write $\mathbb{C}[j,j_j;m]$ in the following very compact form.~\cite{fnote_CG}
\begin{eqnarray}
\mathbb{C}[j,m_j;m] = (-)^{m+|m_j|-j} ~ C^{j}_{|m_j|}C^{m+|m_j|}_j \times \nonumber\\
\frac{\sqrt{C^{2S}_{m+|m_j|}}}{\sqrt{ C^{2S+j+1}_{2j+1} C^{2j}_{j+|m_j|}   C^{2S}_m }} ~ {_3F_2}[a,b,1] \label{eq:compact_CG}
\end{eqnarray}
where ${_3F_2}[a,b,1]$ is the generalized Hypergeometric function, and $a$ and $b$ (two arrays of size 3 and 2, respectively) are given by: $a=\{-j,-j+|m_j|,m+|m_j|-2S\}$ and $b=\{1+|m_j|,1-j+m+|m_j|\}$. (For the definition of $_pF_q$, look up in any book on special functions or Mathematica or {\em Google}.)

We can now explicitly write down any total-spin eigenstate for a pair of spin-$S$. For example, the singlet state (denoted as $|s\rangle$) can easily be written as: \[|s\rangle = \frac{1}{\sqrt{2S+1}}\sum_{m=0}^{2S}(-)^m|S-m,-S+m\rangle. \] We can similarly evaluate the triplets and other higher spin states. Next, we generalize our method to the case of unequal spin quantum numbers, $S_1$ and  $S_2$. 

\subsection{Case of general $S_1$ and $S_2$}
$\bullet$ {\bf\em Proposition $2^{*}$.} The generating state $|j;\xi \rangle$, for the total-spin quantum number $j$, is given by: 
\begin{eqnarray}
 |j;\xi\rangle &=&\left(\ahat_1^\dag+\xi\bhat_1^\dag\right)^{j+S_1-S_2} \left(\ahat_2^\dag+\xi\bhat_2^\dag\right)^{j+S_2-S_1} \nonumber \\
  && \times \left(\calA^\dag\right)^{S_1+S_2-j}|0,0\rangle \nonumber
\end{eqnarray}
where $|S_1-S_2| \le j\le S_1+S_2$.

{\em Proof.} This is a generalization of Proposition 2 (hence, $2^*$). Likewise, evaluate $(\S_1+\S_2)^2|j;\xi\rangle$. We find that 
\begin{eqnarray}
&&(\S_1+\S_2)^2|j;\xi\rangle \nonumber\\
 &&=  j(j+1)|j;\xi\rangle-\calA^{S_1+S_2-j+1}\calA\left[\calB^\dag(\xi)\right]^j|0,0\rangle \nonumber
\end{eqnarray}
Following the same steps as in Proposition 2, we can show that $\calA^{S_1+S_2-j+1}\calA\left[\calB^\dag(\xi)\right]^j|0,0\rangle=0$. Therefore, $(\S_1+\S_2)^2|j;\xi\rangle = j(j+1)|j;\xi\rangle $. 

The range of $j$ is fixed by demanding that the powers of $(\ahat_1^\dag+\xi\bhat_1^\dag)$,  $(\ahat_2^\dag+\xi\bhat_2^\dag)$ and $\calA^\dag$ in the state $|j;\xi\rangle$ must be positive integers. It implies that $j\ge |S_1-S_2|$ and $j\le S_1+S_2$. Physically, the lower bound, $|S_1-S_2|$, is tied to the fact that a maximum of $\mbox{min}(2S_1,2S_2)$ valence bonds can be made between two spins. The upper bound on $j$ is fixed by the total number of Schwinger bosons, $2(S_1+S_2)$. Hence, the proof. $\bullet$.

$\bullet$ {\bf\em Proposition $3^*$.} Series expansion of $|j;\xi\rangle$.
 \[|j;\xi\rangle=\sum_{m_j=j}^{-j}\xi^{j-m_j}|\widetilde{j,m_j}\rangle\]
Here, $|\widetilde{j,m_j}\rangle$ is an unnormalized eigenstate of $(\S_1+\S_2)^2$.

{\em Proof.} Consider $(\ahat_1^\dag+\xi\bhat_1^\dag)^{j+S_1-S_2} (\ahat_2^\dag+\xi\bhat_2^\dag)^{j+S_2-S_1}$ first. It can be expanded as:
\begin{eqnarray}
&&  (\ahat_1^\dag+\xi\bhat_1^\dag)^{j+S_1-S_2} (\ahat_2^\dag+\xi\bhat_2^\dag)^{j+S_2-S_1}  \nonumber\\
 &&~~~ = \sum_{m_j=j}^{-j} \xi^{j-m_j}\calB^\dag(j,m_j) \nonumber
\end{eqnarray}
where 
\begin{eqnarray}
&&\calB^\dag(j,m_j)=  \sum_{l=\mbox{max}(0,S_2-S_1-m_j)}^{j+\mbox{min}(S_2-S_1,-m_j)} C^{j+S_1-S_2}_{j-m_j-l} C^{j+S_2-S_1}_{l} \times  \nonumber \\
&& (\ahat^\dag_1)^{m_j+l+S_1-S_2} (\bhat^\dag_1)^{j-m_j-l} (\ahat^\dag_2)^{j+S_2-S_1-l} (\bhat^\dag_2)^l \label{eq:Bdag_jmj_general}
\end{eqnarray}
The above equation is a generalized version of Eq.~(\ref{eq:Bdag_jmj}). Now it's obvious that $|j;\xi\rangle=\sum_{m_j=j}^{-j}\xi^{j-m_j}|\widetilde{j,m_j}\rangle$, where 
\[|\widetilde{j,m_j}\rangle = \calB^\dag(j,m_j)\left(\calA^\dag\right)^{S_1+S_2-j}|0,0\rangle.\]
The above result, together with Proposition $2^*$, further implies that $(\S_1+\S_2)^2|\widetilde{j,m_j}\rangle=j(j+1)|\widetilde{j,m_j}\rangle$. Moreover, Eq.~(\ref{eq:Bdag_jmj_general}) implies $(\S_1+\S_2)_z|\widetilde{j,m_j}\rangle=m_j|\widetilde{j,m_j}\rangle$ because  $\calA^\dag$ contributes nothing to $(\S_1+\S_2)_z$. $\bullet$

$\bullet$ {\bf\em Proposition $4^*$.} Normalized total-spin eigenstate:
\[ |j,m_j\rangle = \frac{\sqrt{C^{2S_1}_{S_1+S_2-j} C^{2S_2}_{S_1+S_2-j}}}{\sqrt{(2S_1)! \, (2S_2)!  \, C^{S_1+S_2+j+1}_{2j+1} C^{2j}_{j+m_j}}} |\widetilde{j,m_j}\rangle \]
{\em Proof.} Calculate $\langle j;\xi | j;\xi \rangle$, as in Proposition 4. We find,
\begin{eqnarray}
&& \langle j;\xi | j;\xi \rangle = C^{S_1+S_2+j+1}_{2j+1}\left[(S_1+S_2-j)!\right ]^2\times ~~~~~\nonumber \\
&& ~~~~ (j+S_1-S_2)! ~(j+S_2-S_1)!~(1+\xi^2)^{2j}. \nonumber
\end{eqnarray}
Moreover,  $\langle j;\xi | j;\xi \rangle=\sum_{m_j=j}^{-j}\xi^{2(j-m_j)}  \langle\widetilde{j,m_j}|\widetilde{j,m_j}\rangle$, deduced from Proposition $3^*$.
Therefore, 
\begin{eqnarray}
&& \langle\widetilde{j,m_j}|\widetilde{j,m_j}\rangle = C^{S_1+S_2+j+1}_{2j+1} C^{2j}_{j+m_j} \times \nonumber \\
&&~~~ \left[(S_1+S_2-j)!\right ]^2 (j+S_1-S_2)! ~(j+S_2-S_1)! \nonumber \\
&& =\frac{(2S_1)! \, (2S_2)! \, C^{S_1+S_2+j+1}_{2j+1} C^{2j}_{j+m_j}}{ C^{2S_1}_{S_1+S_2-j} C^{2S_2}_{S_1+S_2-j}} \nonumber
\end{eqnarray}
Hence, the proof. $\bullet$

$\bullet$ {\bf\em Proposition $5^*$.} The Clebsch-Gordan coefficients.
\[ |j,m_j\rangle = \sum_{m=m_{min}}^{m_{max}}\mathbb{C}[j,m_j;m]|S_1-m,-S_1+m+m_j\rangle \]
where 
\begin{eqnarray}
m_{min} &=& -\mbox{min}(0,m_j+S_2-S_1), \nonumber \\
m_{max} &=& S_1+S_2-\mbox{max}(m_j, S_2-S_1), \nonumber
\end{eqnarray}
and the Clebsch-Gordan coefficients, $\mathbb{C}[j,m_j;m]$, are given by
\begin{eqnarray}
&& \mathbb{C}[j,m_j;m] \nonumber \\
&& ~~= \frac{ \sqrt{C^{2S_1}_{S_1+S_2-j} C^{2S_2}_{S_1+S_2-j}} }{\sqrt{C^{S_1+S_2+j+1}_{2j+1} C^{2j}_{j+m_j} C^{2S_2}_{m+m_j+S_2-S_1} C^{2S_1}_{m} } } \nonumber \\
&& ~~\times \sum_{p=p_{min}}^{p_{max}} (-)^p C^{j+S_1-S_2}_{m-p} C^{j+S_2-S_1}_{j-m_j-m+p} C^{S_1+S_2-j}_p \nonumber
\end{eqnarray}
where 
\begin{eqnarray}
p_{min} &=& \mbox{max}[0,m+\mbox{max}(m_j,S_2-S_1)-j],~\mbox{and} \nonumber\\
p_{max} &=& \mbox{min}[S_1+S_2-j,m+\mbox{min}(0,m_j+S_2-S_1)].\nonumber
\end{eqnarray}

{\em Proof.} One can get it from Propostions $3^*$ and $4^*$, by carefully doing a few steps of algebra, similar to that in Proposition 5. Note that this proposition correctly reproduces Proposition 5 for $S_1=S_2=S$.$\bullet$

This completes our description of a Schwinger-boson based method of constructing the total-spin eigenstates for a pair of quantum spins.

\bibliography{references}

\begin{thebibliography}{10}%
\makeatletter
\providecommand \@ifxundefined [1]{%
 \ifx #1\undefined \expandafter \@firstoftwo
 \else \expandafter \@secondoftwo
\fi
}%
\providecommand \@ifnum [1]{%
 \ifnum #1\expandafter \@firstoftwo
 \else \expandafter \@secondoftwo
\fi
}%
\providecommand \enquote [1]{``#1''}%
\providecommand \bibnamefont  [1]{#1}%
\providecommand \bibfnamefont [1]{#1}%
\providecommand \citenamefont [1]{#1}%
\providecommand\href[0]{\@sanitize\@href}%
\providecommand\@href[1]{\endgroup\@@startlink{#1}\endgroup\@@href}%
\providecommand\@@href[1]{#1\@@endlink}%
\providecommand \@sanitize [0]{\begingroup\catcode`\&12\catcode`\#12\relax}%
\@ifxundefined \pdfoutput {\@firstoftwo}{%
 \@ifnum{\z@=\pdfoutput}{\@firstoftwo}{\@secondoftwo}%
}{%
 \providecommand\@@startlink[1]{\leavevmode\special{html:<a href="#1">}}%
 \providecommand\@@endlink[0]{\special{html:</a>}}%
}{%
 \providecommand\@@startlink[1]{%
  \leavevmode
  \pdfstartlink
   attr{/Border[0 0 1 ]/H/I/C[0 1 1]}%
   user{/Subtype/Link/A<</Type/Action/S/URI/URI(#1)>>}%
  \relax
 }%
 \providecommand\@@endlink[0]{\pdfendlink}%
}%
\providecommand \url  [0]{\begingroup\@sanitize \@url }%
\providecommand \@url [1]{\endgroup\@href {#1}{\urlprefix}}%
\providecommand \urlprefix [0]{URL }%
\providecommand \Eprint[0]{\href }%
\@ifxundefined \urlstyle {%
  \providecommand \doi [1]{doi:\discretionary{}{}{}#1}%
}{%
  \providecommand \doi [0]{doi:\discretionary{}{}{}\begingroup
  \urlstyle{rm}\Url }%
}%
\providecommand \doibase [0]{http://dx.doi.org/}%
\providecommand \Doi[1]{\href{\doibase#1}}%
\providecommand \bibAnnote [3]{%
  \BibitemShut{#1}%
  \begin{quotation}\noindent
    \textsc{Key:}\ #2\\\textsc{Annotation:}\ #3%
  \end{quotation}%
}%
\providecommand \bibAnnoteFile [2]{%
  \IfFileExists{#2}{\bibAnnote {#1} {#2} {\input{#2}}}{}%
}%
\providecommand \typeout [0]{\immediate \write \m@ne }%
\providecommand \selectlanguage [0]{\@gobble}%
\providecommand \bibinfo [0]{\@secondoftwo}%
\providecommand \bibfield [0]{\@secondoftwo}%
\providecommand \translation [1]{[#1]}%
\providecommand \BibitemOpen[0]{}%
\providecommand \bibitemStop [0]{}%
\providecommand \bibitemNoStop [0]{.\EOS\space}%
\providecommand \EOS [0]{\spacefactor3000\relax}%
\providecommand \BibitemShut [1]{\csname bibitem#1\endcsname}%
\bibitem{Lhuillier}%
  \BibitemOpen
  \bibfield{author}{%
  \bibinfo {author} {\bibfnamefont{G.}~\bibnamefont{Misguich}}\ and\ \bibinfo
  {author} {\bibfnamefont{C.}~\bibnamefont{Lhuillier}},\ }%
  in\ \emph{\bibinfo {booktitle} {Frustrated Spin Systems}},\ \bibinfo {series
  and number} {\bibinfo {number} {arXiv:cond-mat/0310405}},\ \bibinfo {editor}
  {edited by\ \bibinfo {editor} {\bibfnamefont{H.~T.}\ \bibnamefont{Diep}}}\
  (\bibinfo {publisher} {World-Scientific, Singapore},\ \bibinfo {year}
  {2005})%
  \bibAnnoteFile{NoStop}{Lhuillier}%
\bibitem{Indrani}%
  \BibitemOpen
  \bibfield{author}{%
  \bibinfo {author} {\bibfnamefont{I.}~\bibnamefont{Bose}},\ }%
  in\ \emph{\bibinfo {booktitle} {Field Theories in Condensed Matter
  Physics}},\ \bibinfo {series and number} {\bibinfo {number}
  {arXiv:cond-mat/0011262}},\ \bibinfo {editor} {edited by\ \bibinfo {editor}
  {\bibfnamefont{S.}~\bibnamefont{Rao}}}\ (\bibinfo {publisher} {Hindustan Book
  Agency},\ \bibinfo {year} {2000})%
  \bibAnnoteFile{NoStop}{Indrani}%
\bibitem{MG}%
  \BibitemOpen
  \bibfield{author}{%
  \bibinfo {author} {\bibfnamefont{C.~K.}\ \bibnamefont{Majumdar}}\ and\
  \bibinfo {author} {\bibfnamefont{D.~K.}\ \bibnamefont{Ghosh}},\ }%
  \bibfield{journal}{%
  \bibinfo {journal} {J. Math. Phys. (N.Y.)}\ }%
  \textbf{\bibinfo {volume} {10}},\ \bibinfo {pages} {1399} (\bibinfo {year}
  {1969})%
  \bibAnnoteFile{NoStop}{MG}%
\bibitem{SS}%
  \BibitemOpen
  \bibfield{author}{%
  \bibinfo {author} {\bibfnamefont{B.~S.}\ \bibnamefont{Shastry}}\ and\
  \bibinfo {author} {\bibfnamefont{B.}~\bibnamefont{Sutherland}},\ }%
  \bibfield{journal}{%
  \bibinfo {journal} {Physica B \& C}\ }%
  \textbf{\bibinfo {volume} {108}},\ \bibinfo {pages} {1069} (\bibinfo {year}
  {1981})%
  \bibAnnoteFile{NoStop}{SS}%
\bibitem{Kageyama}%
  \BibitemOpen
  \bibfield{author}{%
  \bibinfo {author} {\bibfnamefont{H.}~\bibnamefont{Kageyama}}, \bibinfo
  {author} {\bibfnamefont{K.}~\bibnamefont{Yoshimura}}, \bibinfo {author}
  {\bibfnamefont{R.}~\bibnamefont{Stern}}, \bibinfo {author}
  {\bibfnamefont{N.~V.}\ \bibnamefont{Mushnikov}}, \bibinfo {author}
  {\bibfnamefont{K.}~\bibnamefont{Onizuka}}, \bibinfo {author}
  {\bibfnamefont{M.}~\bibnamefont{Kato}}, \bibinfo {author}
  {\bibfnamefont{K.}~\bibnamefont{Kosuge}}, \bibinfo {author}
  {\bibfnamefont{C.~P.}\ \bibnamefont{Slichter}}, \bibinfo {author}
  {\bibfnamefont{T.}~\bibnamefont{Goto}},\ and\ \bibinfo {author}
  {\bibfnamefont{Y.}~\bibnamefont{Ueda}},\ }%
  \bibfield{journal}{%
  \bibinfo {journal} {Phys. Rev. Lett.}\ }%
  \textbf{\bibinfo {volume} {82}},\ \bibinfo {pages} {3168} (\bibinfo {year}
  {1999})%
  \bibAnnoteFile{NoStop}{Kageyama}%
\bibitem{CaV4O9}%
  \BibitemOpen
  \bibfield{author}{%
  \bibinfo {author} {\bibfnamefont{S.}~\bibnamefont{Taniguchi}}, \bibinfo
  {author} {\bibfnamefont{T.}~\bibnamefont{Nishikawa}}, \bibinfo {author}
  {\bibfnamefont{Y.}~\bibnamefont{Yasui}}, \bibinfo {author}
  {\bibfnamefont{Y.}~\bibnamefont{Kobayashi}}, \bibinfo {author}
  {\bibfnamefont{M.}~\bibnamefont{Sato}}, \bibinfo {author}
  {\bibfnamefont{T.}~\bibnamefont{Nishioka}}, \bibinfo {author}
  {\bibfnamefont{M.}~\bibnamefont{Kontani}},\ and\ \bibinfo {author}
  {\bibfnamefont{K.}~\bibnamefont{Sano}},\ }%
  \bibfield{journal}{%
  \bibinfo {journal} {J. Phys. Soc. Jpn.}\ }%
  \textbf{\bibinfo {volume} {64}},\ \bibinfo {pages} {2758} (\bibinfo {year}
  {1995})%
  \bibAnnoteFile{NoStop}{CaV4O9}%
\bibitem{CaV4O9_Troyer}%
  \BibitemOpen
  \bibfield{author}{%
  \bibinfo {author} {\bibfnamefont{M.}~\bibnamefont{Troyer}}, \bibinfo {author}
  {\bibfnamefont{H.}~\bibnamefont{Kontani}},\ and\ \bibinfo {author}
  {\bibfnamefont{K.}~\bibnamefont{Ueda}},\ }%
  \bibfield{journal}{%
  \bibinfo {journal} {Phys. Rev. Lett.}\ }%
  \textbf{\bibinfo {volume} {76}},\ \bibinfo {pages} {3822} (\bibinfo {year}
  {1996})%
  \bibAnnoteFile{NoStop}{CaV4O9_Troyer}%
\bibitem{Kagome_Material}%
  \BibitemOpen
  \bibfield{author}{%
  \bibinfo {author} {\bibfnamefont{P.}~\bibnamefont{Mendels}}, \bibinfo
  {author} {\bibfnamefont{F.}~\bibnamefont{Bert}}, \bibinfo {author}
  {\bibfnamefont{M.~A.}\ \bibnamefont{de~Vries}}, \bibinfo {author}
  {\bibfnamefont{A.}~\bibnamefont{Olariu}}, \bibinfo {author}
  {\bibfnamefont{A.}~\bibnamefont{Harrison}}, \bibinfo {author}
  {\bibfnamefont{F.}~\bibnamefont{Duc}}, \bibinfo {author}
  {\bibfnamefont{J.~C.}\ \bibnamefont{Trombe}}, \bibinfo {author}
  {\bibfnamefont{J.~S.}\ \bibnamefont{Lord}}, \bibinfo {author}
  {\bibfnamefont{A.}~\bibnamefont{Amato}},\ and\ \bibinfo {author}
  {\bibfnamefont{C.}~\bibnamefont{Baines}},\ }%
  \bibfield{journal}{%
  \bibinfo {journal} {Phys. Rev. Lett.}\ }%
  \textbf{\bibinfo {volume} {98}},\ \bibinfo {pages} {077204} (\bibinfo {year}
  {2007})%
  \bibAnnoteFile{NoStop}{Kagome_Material}%
\bibitem{Mila}%
  \BibitemOpen
  \bibfield{author}{%
  \bibinfo {author} {\bibfnamefont{F.}~\bibnamefont{Mila}},\ }%
  \bibfield{journal}{%
  \bibinfo {journal} {Phys. Rev. Lett.}\ }%
  \textbf{\bibinfo {volume} {81}},\ \bibinfo {pages} {2356} (\bibinfo {year}
  {1998})%
  \bibAnnoteFile{NoStop}{Mila}%
\bibitem{kappa_salt}%
  \BibitemOpen
  \bibfield{author}{%
  \bibinfo {author} {\bibfnamefont{S.}~\bibnamefont{Yamashita}}, \bibinfo
  {author} {\bibfnamefont{Y.}~\bibnamefont{Nakazawa}}, \bibinfo {author}
  {\bibfnamefont{M.}~\bibnamefont{Oguni}}, \bibinfo {author}
  {\bibfnamefont{Y.}~\bibnamefont{Oshima}}, \bibinfo {author}
  {\bibfnamefont{H.}~\bibnamefont{Nojiri}}, \bibinfo {author}
  {\bibfnamefont{Y.}~\bibnamefont{Shimizu}}, \bibinfo {author}
  {\bibfnamefont{K.}~\bibnamefont{Miyagawa}},\ and\ \bibinfo {author}
  {\bibfnamefont{K.}~\bibnamefont{Kanoda}},\ }%
  \bibfield{journal}{%
  \bibinfo {journal} {Nature Physics}\ }%
  \textbf{\bibinfo {volume} {4}},\ \bibinfo {pages} {459} (\bibinfo {year}
  {2008})%
  \bibAnnoteFile{NoStop}{kappa_salt}%
\bibitem{TlCuCl3}%
  \BibitemOpen
  \bibfield{author}{%
  \bibinfo {author} {\bibfnamefont{A.}~\bibnamefont{Oosawa}}, \bibinfo {author}
  {\bibfnamefont{H.}~\bibnamefont{Ishii}},\ and\ \bibinfo {author}
  {\bibfnamefont{H.}~\bibnamefont{Tanaka}},\ }%
  \bibfield{journal}{%
  \bibinfo {journal} {J. Phys.: Condens. Matter}\ }%
  \textbf{\bibinfo {volume} {11}},\ \bibinfo {pages} {265} (\bibinfo {year}
  {1999})%
  \bibAnnoteFile{NoStop}{TlCuCl3}%
\bibitem{TlCuCl3Nature}%
  \BibitemOpen
  \bibfield{author}{%
  \bibinfo {author} {\bibfnamefont{C.}~\bibnamefont{R\"uegg}}, \bibinfo
  {author} {\bibfnamefont{N.}~\bibnamefont{Cavadini}}, \bibinfo {author}
  {\bibfnamefont{A.}~\bibnamefont{Furrer}}, \bibinfo {author}
  {\bibfnamefont{H.-U.}\ \bibnamefont{G\"udel}}, \bibinfo {author}
  {\bibfnamefont{K.}~\bibnamefont{Kr\"amer}}, \bibinfo {author}
  {\bibfnamefont{H.}~\bibnamefont{Mutka}}, \bibinfo {author}
  {\bibfnamefont{A.}~\bibnamefont{Wildes}}, \bibinfo {author}
  {\bibfnamefont{K.}~\bibnamefont{Habicht}},\ and\ \bibinfo {author}
  {\bibfnamefont{P.}~\bibnamefont{Vorderwisch}},\ }%
  \bibfield{journal}{%
  \bibinfo {journal} {Nature (London)}\ }%
  \textbf{\bibinfo {volume} {423}},\ \bibinfo {pages} {62} (\bibinfo {year}
  {2003})%
  \bibAnnoteFile{NoStop}{TlCuCl3Nature}%
\bibitem{BaCuSi2O6}%
  \BibitemOpen
  \bibfield{author}{%
  \bibinfo {author} {\bibfnamefont{Y.}~\bibnamefont{Sasago}}, \bibinfo {author}
  {\bibfnamefont{K.}~\bibnamefont{Uchinokura}}, \bibinfo {author}
  {\bibfnamefont{A.}~\bibnamefont{Zheludev}},\ and\ \bibinfo {author}
  {\bibfnamefont{G.}~\bibnamefont{Shirane}},\ }%
  \bibfield{journal}{%
  \bibinfo {journal} {Phys. Rev. B}\ }%
  \textbf{\bibinfo {volume} {55}},\ \bibinfo {pages} {8357} (\bibinfo {year}
  {1997})%
  \bibAnnoteFile{NoStop}{BaCuSi2O6}%
\bibitem{CuFeGeO}%
  \BibitemOpen
  \bibfield{author}{%
  \bibinfo {author} {\bibfnamefont{T.}~\bibnamefont{Masuda}}, \bibinfo {author}
  {\bibfnamefont{A.}~\bibnamefont{Zheludev}}, \bibinfo {author}
  {\bibfnamefont{B.}~\bibnamefont{Sales}}, \bibinfo {author}
  {\bibfnamefont{S.}~\bibnamefont{Imai}}, \bibinfo {author}
  {\bibfnamefont{K.}~\bibnamefont{Uchinokura}},\ and\ \bibinfo {author}
  {\bibfnamefont{S.}~\bibnamefont{Park}},\ }%
  \bibfield{journal}{%
  \bibinfo {journal} {Phys. Rev. B}\ }%
  \textbf{\bibinfo {volume} {72}},\ \bibinfo {pages} {094434} (\bibinfo {year}
  {2005})%
  \bibAnnoteFile{NoStop}{CuFeGeO}%
\bibitem{Sr3Cr2O8}%
  \BibitemOpen
  \bibfield{author}{%
  \bibinfo {author} {\bibfnamefont{D.~L.}\ \bibnamefont{Quintero-Castro}},
  \bibinfo {author} {\bibfnamefont{B.}~\bibnamefont{Lake}}, \bibinfo {author}
  {\bibfnamefont{E.~M.}\ \bibnamefont{Wheeler}}, \bibinfo {author}
  {\bibfnamefont{A.~T. M.~N.}\ \bibnamefont{Islam}}, \bibinfo {author}
  {\bibfnamefont{T.}~\bibnamefont{Guidi}}, \bibinfo {author}
  {\bibfnamefont{K.~C.}\ \bibnamefont{Rule}}, \bibinfo {author}
  {\bibfnamefont{Z.}~\bibnamefont{Izaola}}, \bibinfo {author}
  {\bibfnamefont{M.}~\bibnamefont{Russina}}, \bibinfo {author}
  {\bibfnamefont{K.}~\bibnamefont{Kiefer}},\ and\ \bibinfo {author}
  {\bibfnamefont{Y.}~\bibnamefont{Skourski}},\ }%
  \bibfield{journal}{%
  \bibinfo {journal} {Phys. Rev. B}\ }%
  \textbf{\bibinfo {volume} {81}},\ \bibinfo {pages} {014415} (\bibinfo {year}
  {2010})%
  \bibAnnoteFile{NoStop}{Sr3Cr2O8}%
\bibitem{sach-bhatt}%
  \BibitemOpen
  \bibfield{author}{%
  \bibinfo {author} {\bibfnamefont{S.}~\bibnamefont{Sachdev}}\ and\ \bibinfo
  {author} {\bibfnamefont{R.~N.}\ \bibnamefont{Bhatt}},\ }%
  \bibfield{journal}{%
  \bibinfo {journal} {Phys. Rev. B}\ }%
  \textbf{\bibinfo {volume} {41}},\ \bibinfo {pages} {9323} (\bibinfo {year}
  {1990})%
  \bibAnnoteFile{NoStop}{sach-bhatt}%
\bibitem{chub}%
  \BibitemOpen
  \bibfield{author}{%
  \bibinfo {author} {\bibfnamefont{A.~V.}\ \bibnamefont{Chubukov}},\ }%
  \bibfield{journal}{%
  \bibinfo {journal} {JETP Lett.}\ }%
  \textbf{\bibinfo {volume} {49}},\ \bibinfo {pages} {129} (\bibinfo {year}
  {1989})%
  \bibAnnoteFile{NoStop}{chub}%
\bibitem{gopalan_rice}%
  \BibitemOpen
  \bibfield{author}{%
  \bibinfo {author} {\bibfnamefont{S.}~\bibnamefont{Gopalan}}, \bibinfo
  {author} {\bibfnamefont{T.~M.}\ \bibnamefont{Rice}},\ and\ \bibinfo {author}
  {\bibfnamefont{M.}~\bibnamefont{Sigrist}},\ }%
  \bibfield{journal}{%
  \bibinfo {journal} {Phys. Rev. B}\ }%
  \textbf{\bibinfo {volume} {49}},\ \bibinfo {pages} {8901} (\bibinfo {year}
  {1994})%
  \bibAnnoteFile{NoStop}{gopalan_rice}%
\bibitem{rkbk}%
  \BibitemOpen
  \bibfield{author}{%
  \bibinfo {author} {\bibfnamefont{R.}~\bibnamefont{Kumar}}\ and\ \bibinfo
  {author} {\bibfnamefont{B.}~\bibnamefont{Kumar}},\ }%
  \bibfield{journal}{%
  \bibinfo {journal} {Phys. Rev. B}\ }%
  \textbf{\bibinfo {volume} {77}},\ \bibinfo {pages} {144413} (\bibinfo {year}
  {2008})%
  \bibAnnoteFile{NoStop}{rkbk}%
\bibitem{rkdkbk}%
  \BibitemOpen
  \bibfield{author}{%
  \bibinfo {author} {\bibfnamefont{R.}~\bibnamefont{Kumar}}, \bibinfo {author}
  {\bibfnamefont{D.}~\bibnamefont{Kumar}},\ and\ \bibinfo {author}
  {\bibfnamefont{B.}~\bibnamefont{Kumar}},\ }%
  \bibfield{journal}{%
  \bibinfo {journal} {Phys. Rev. B}\ }%
  \textbf{\bibinfo {volume} {80}},\ \bibinfo {pages} {214428} (\bibinfo {year}
  {2009})%
  \bibAnnoteFile{NoStop}{rkdkbk}%
\bibitem{spin1_bo1}%
  \BibitemOpen
  \bibfield{author}{%
  \bibinfo {author} {\bibfnamefont{H.-T.}\ \bibnamefont{Wang}}, \bibinfo
  {author} {\bibfnamefont{H.~Q.}\ \bibnamefont{Lin}},\ and\ \bibinfo {author}
  {\bibfnamefont{J.-L.}\ \bibnamefont{Shen}},\ }%
  \bibfield{journal}{%
  \bibinfo {journal} {Phys. Rev. B}\ }%
  \textbf{\bibinfo {volume} {61}},\ \bibinfo {pages} {4019} (\bibinfo {year}
  {2000})%
  \bibAnnoteFile{NoStop}{spin1_bo1}%
\bibitem{spin1_bo2}%
  \BibitemOpen
  \bibfield{author}{%
  \bibinfo {author} {\bibfnamefont{B.}~\bibnamefont{Xu}}, \bibinfo {author}
  {\bibfnamefont{H.-T.}\ \bibnamefont{Wang}},\ and\ \bibinfo {author}
  {\bibfnamefont{Y.}~\bibnamefont{Wang}},\ }%
  \bibfield{journal}{%
  \bibinfo {journal} {Phys. Rev. B}\ }%
  \textbf{\bibinfo {volume} {77}},\ \bibinfo {pages} {014401} (\bibinfo {year}
  {2008})%
  \bibAnnoteFile{NoStop}{spin1_bo2}%
\bibitem{spin1_bo3}%
  \BibitemOpen
  \bibfield{author}{%
  \bibinfo {author} {\bibfnamefont{A.~V.}\ \bibnamefont{Chubukov}},\ }%
  \bibfield{journal}{%
  \bibinfo {journal} {Phys. Rev. B}\ }%
  \textbf{\bibinfo {volume} {43}},\ \bibinfo {pages} {3337} (\bibinfo {year}
  {1991})%
  \bibAnnoteFile{NoStop}{spin1_bo3}%
\bibitem{spin1_bo4}%
  \BibitemOpen
  \bibfield{author}{%
  \bibinfo {author} {\bibfnamefont{W.}~\bibnamefont{Brenig}}\ and\ \bibinfo
  {author} {\bibfnamefont{K.~W.}\ \bibnamefont{Becker}},\ }%
  \bibfield{journal}{%
  \bibinfo {journal} {Phys. Rev. B}\ }%
  \textbf{\bibinfo {volume} {64}},\ \bibinfo {pages} {214413} (\bibinfo {year}
  {2001})%
  \bibAnnoteFile{NoStop}{spin1_bo4}%
\bibitem{plaquette1}%
  \BibitemOpen
  \bibfield{author}{%
  \bibinfo {author} {\bibfnamefont{M.~E.}\ \bibnamefont{Zhitomirsky}}\ and\
  \bibinfo {author} {\bibfnamefont{K.}~\bibnamefont{Ueda}},\ }%
  \bibfield{journal}{%
  \bibinfo {journal} {Phys. Rev. B}\ }%
  \textbf{\bibinfo {volume} {54}},\ \bibinfo {pages} {9007} (\bibinfo {year}
  {1996})%
  \bibAnnoteFile{NoStop}{plaquette1}%
\bibitem{Schwinger}%
  \BibitemOpen
  \bibfield{author}{%
  \bibinfo {author} {\bibfnamefont{J.}~\bibnamefont{Schwinger}},\ }%
  in\ \emph{\bibinfo {booktitle} {Quantum Theory of Angular Momentum}},\
  \bibinfo {editor} {edited by\ \bibinfo {editor} {\bibfnamefont{L.~C.}\
  \bibnamefont{Biedenharn}}\ and\ \bibinfo {editor}
  {\bibfnamefont{H.}~\bibnamefont{van Dam}}}\ (\bibinfo {publisher} {Academic
  Press},\ \bibinfo {year} {1965})%
  \bibAnnoteFile{NoStop}{Schwinger}%
\bibitem{Mattis}%
  \BibitemOpen
  \bibfield{author}{%
  \bibinfo {author} {\bibfnamefont{D.~C.}\ \bibnamefont{Mattis}},\ }%
  \emph{\bibinfo {title} {The Theory of Magnetism Made Simple}}\ (\bibinfo
  {publisher} {World-Scientific},\ \bibinfo {year} {2006})%
  \bibAnnoteFile{NoStop}{Mattis}%
\bibitem{SinghGelfandHuse}%
  \BibitemOpen
  \bibfield{author}{%
  \bibinfo {author} {\bibfnamefont{R.~R.~P.}\ \bibnamefont{Singh}}, \bibinfo
  {author} {\bibfnamefont{M.~P.}\ \bibnamefont{Gelfand}},\ and\ \bibinfo
  {author} {\bibfnamefont{D.~A.}\ \bibnamefont{Huse}},\ }%
  \bibfield{journal}{%
  \bibinfo {journal} {Phys. Rev. Lett.}\ }%
  \textbf{\bibinfo {volume} {61}},\ \bibinfo {pages} {2484} (\bibinfo {year}
  {1988})%
  \bibAnnoteFile{NoStop}{SinghGelfandHuse}%
\bibitem{AffleckGelfandSingh}%
  \BibitemOpen
  \bibfield{author}{%
  \bibinfo {author} {\bibfnamefont{I.}~\bibnamefont{Affleck}}, \bibinfo
  {author} {\bibfnamefont{M.~P.}\ \bibnamefont{Gelfand}},\ and\ \bibinfo
  {author} {\bibfnamefont{R.~R.~P.}\ \bibnamefont{Singh}},\ }%
  \bibfield{journal}{%
  \bibinfo {journal} {J. Phys. A: Math. Gen.}\ }%
  \textbf{\bibinfo {volume} {27}},\ \bibinfo {pages} {7313} (\bibinfo {year}
  {1994})%
  \bibAnnoteFile{NoStop}{AffleckGelfandSingh}%
\bibitem{Matsumoto}%
  \BibitemOpen
  \bibfield{author}{%
  \bibinfo {author} {\bibfnamefont{M.}~\bibnamefont{Matsumoto}}, \bibinfo
  {author} {\bibfnamefont{C.}~\bibnamefont{Yasuda}}, \bibinfo {author}
  {\bibfnamefont{S.}~\bibnamefont{Todo}},\ and\ \bibinfo {author}
  {\bibfnamefont{H.}~\bibnamefont{Takayama}},\ }%
  \bibfield{journal}{%
  \bibinfo {journal} {Phys. Rev. B}\ }%
  \textbf{\bibinfo {volume} {65}},\ \bibinfo {pages} {014407} (\bibinfo {year}
  {2001})%
  \bibAnnoteFile{NoStop}{Matsumoto}%
\bibitem{Ivanov}%
  \BibitemOpen
  \bibfield{author}{%
  \bibinfo {author} {\bibfnamefont{N.~B.}\ \bibnamefont{Ivanov}}, \bibinfo
  {author} {\bibfnamefont{S.~E.}\ \bibnamefont{Kr\"uger}},\ and\ \bibinfo
  {author} {\bibfnamefont{J.}~\bibnamefont{Richter}},\ }%
  \bibfield{journal}{%
  \bibinfo {journal} {Phys. Rev. B}\ }%
  \textbf{\bibinfo {volume} {53}},\ \bibinfo {pages} {2633} (\bibinfo {year}
  {1996})%
  \bibAnnoteFile{NoStop}{Ivanov}%
\bibitem{Imada}%
  \BibitemOpen
  \bibfield{author}{%
  \bibinfo {author} {\bibfnamefont{N.}~\bibnamefont{Katoh}}\ and\ \bibinfo
  {author} {\bibfnamefont{M.}~\bibnamefont{Imada}},\ }%
  \bibfield{journal}{%
  \bibinfo {journal} {J. Phys. Soc. Jpn.}\ }%
  \textbf{\bibinfo {volume} {62}},\ \bibinfo {pages} {3728} (\bibinfo {year}
  {1993})%
  \bibAnnoteFile{NoStop}{Imada}%
\bibitem{Richter_SpinDependence}%
  \BibitemOpen
  \bibfield{author}{%
  \bibinfo {author} {\bibfnamefont{R.}~\bibnamefont{Darradi}}, \bibinfo
  {author} {\bibfnamefont{J.}~\bibnamefont{Richter}},\ and\ \bibinfo {author}
  {\bibfnamefont{D.~J.~J.}\ \bibnamefont{Farnell}},\ }%
  \bibfield{journal}{%
  \bibinfo {journal} {J. Phys.: Condens. Matter}\ }%
  \textbf{\bibinfo {volume} {17}},\ \bibinfo {pages} {341} (\bibinfo {year}
  {2005})%
  \bibAnnoteFile{NoStop}{Richter_SpinDependence}%
\bibitem{Wenzel_Janke}%
  \BibitemOpen
  \bibfield{author}{%
  \bibinfo {author} {\bibfnamefont{S.}~\bibnamefont{Wenzel}}\ and\ \bibinfo
  {author} {\bibfnamefont{W.}~\bibnamefont{Janke}},\ }%
  \bibfield{journal}{%
  \bibinfo {journal} {Phys. Rev. B}\ }%
  \textbf{\bibinfo {volume} {79}},\ \bibinfo {pages} {014410} (\bibinfo {year}
  {2009})%
  \bibAnnoteFile{NoStop}{Wenzel_Janke}%
\bibitem{Sandvik}%
  \BibitemOpen
  \bibfield{author}{%
  \bibinfo {author} {\bibfnamefont{A.~W.}\ \bibnamefont{Sandvik}},\ }%
  \bibfield{journal}{%
  \bibinfo {journal} {Phys. Rev. Lett.}\ }%
  \textbf{\bibinfo {volume} {83}},\ \bibinfo {pages} {3069} (\bibinfo {year}
  {1999})%
  \bibAnnoteFile{NoStop}{Sandvik}%
\bibitem{j1j2_Chandra_Doucot}%
  \BibitemOpen
  \bibfield{author}{%
  \bibinfo {author} {\bibfnamefont{P.}~\bibnamefont{Chandra}}\ and\ \bibinfo
  {author} {\bibfnamefont{B.}~\bibnamefont{Doucot}},\ }%
  \bibfield{journal}{%
  \bibinfo {journal} {Phys. Rev. B}\ }%
  \textbf{\bibinfo {volume} {38}},\ \bibinfo {pages} {9335} (\bibinfo {year}
  {1988})%
  \bibAnnoteFile{NoStop}{j1j2_Chandra_Doucot}%
\bibitem{vbc1}%
  \BibitemOpen
  \bibfield{author}{%
  \bibinfo {author} {\bibfnamefont{N.}~\bibnamefont{Read}}\ and\ \bibinfo
  {author} {\bibfnamefont{S.}~\bibnamefont{Sachdev}},\ }%
  \bibfield{journal}{%
  \bibinfo {journal} {Phys. Rev. Lett.}\ }%
  \textbf{\bibinfo {volume} {66}},\ \bibinfo {pages} {1773} (\bibinfo {year}
  {1991})%
  \bibAnnoteFile{NoStop}{vbc1}%
\bibitem{vbc2}%
  \BibitemOpen
  \bibfield{author}{%
  \bibinfo {author} {\bibfnamefont{V.~N.}\ \bibnamefont{Kotov}}, \bibinfo
  {author} {\bibfnamefont{J.}~\bibnamefont{Oitmaa}}, \bibinfo {author}
  {\bibfnamefont{O.~P.}\ \bibnamefont{Sushkov}},\ and\ \bibinfo {author}
  {\bibfnamefont{Z.}~\bibnamefont{Weihong}},\ }%
  \bibfield{journal}{%
  \bibinfo {journal} {Philos. Mag.}\ }%
  \textbf{\bibinfo {volume} {80}},\ \bibinfo {pages} {1483} (\bibinfo {year}
  {2000})%
  \bibAnnoteFile{NoStop}{vbc2}%
\bibitem{vbc4}%
  \BibitemOpen
  \bibfield{author}{%
  \bibinfo {author} {\bibfnamefont{M.~P.}\ \bibnamefont{Gelfand}}, \bibinfo
  {author} {\bibfnamefont{R.~R.~P.}\ \bibnamefont{Singh}},\ and\ \bibinfo
  {author} {\bibfnamefont{D.~A.}\ \bibnamefont{Huse}},\ }%
  \bibfield{journal}{%
  \bibinfo {journal} {Phys. Rev. B}\ }%
  \textbf{\bibinfo {volume} {40}},\ \bibinfo {pages} {10801} (\bibinfo {year}
  {1989})%
  \bibAnnoteFile{NoStop}{vbc4}%
\bibitem{vbc5}%
  \BibitemOpen
  \bibfield{author}{%
  \bibinfo {author} {\bibfnamefont{O.~P.}\ \bibnamefont{Sushkov}}, \bibinfo
  {author} {\bibfnamefont{J.}~\bibnamefont{Oitmaa}},\ and\ \bibinfo {author}
  {\bibfnamefont{Z.}~\bibnamefont{Weihong}},\ }%
  \bibfield{journal}{%
  \bibinfo {journal} {Phys. Rev. B}\ }%
  \textbf{\bibinfo {volume} {63}},\ \bibinfo {pages} {104420} (\bibinfo {year}
  {2001})%
  \bibAnnoteFile{NoStop}{vbc5}%
\bibitem{fnote_AKLT}%
  \BibitemOpen
  \bibinfo {note} {The valence-bond operator, $\calA$, appears prominently in
  the discussion of AKLT states~\cite{AKLT,AKLT2}.}%
  \bibAnnoteFile{Stop}{fnote_AKLT}%
\bibitem{fnote_CG}%
  \BibitemOpen
  \bibinfo {note} {After reflecting upon it for a while, it becomes clear that
  $\sum_{p=\mbox{max}(0,m+|m_j|-j)}^{\mbox{min}(2S-j,m)}$ here in the
  definition of the Clebsch-Gordan coefficients can be written as
  $\sum_{p=m+|m_j|-j}^{m}$.}%
  \bibAnnoteFile{Stop}{fnote_CG}%
\bibitem{AKLT}%
  \BibitemOpen
  \bibfield{author}{%
  \bibinfo {author} {\bibfnamefont{I.}~\bibnamefont{Affleck}}, \bibinfo
  {author} {\bibfnamefont{T.}~\bibnamefont{Kennedy}}, \bibinfo {author}
  {\bibfnamefont{E.~H.}\ \bibnamefont{Lieb}},\ and\ \bibinfo {author}
  {\bibfnamefont{H.}~\bibnamefont{Tasaki}},\ }%
  \bibfield{journal}{%
  \bibinfo {journal} {Phys. Rev. Lett.}\ }%
  \textbf{\bibinfo {volume} {59}},\ \bibinfo {pages} {799} (\bibinfo {year}
  {1987})%
  \bibAnnoteFile{NoStop}{AKLT}%
\bibitem{AKLT2}%
  \BibitemOpen
  \bibfield{author}{%
  \bibinfo {author} {\bibfnamefont{I.}~\bibnamefont{Affleck}}, \bibinfo
  {author} {\bibfnamefont{T.}~\bibnamefont{Kennedy}}, \bibinfo {author}
  {\bibfnamefont{E.~H.}\ \bibnamefont{Lieb}},\ and\ \bibinfo {author}
  {\bibfnamefont{H.}~\bibnamefont{Tasaki}},\ }%
  \bibfield{journal}{%
  \bibinfo {journal} {Commun. Math. Phys.}\ }%
  \textbf{\bibinfo {volume} {115}},\ \bibinfo {pages} {477} (\bibinfo {year}
  {1988})%
  \bibAnnoteFile{NoStop}{AKLT2}%
\end{thebibliography}%
\end{document}